\documentclass[onecolumn, 11pt]{elsarticle}

\usepackage{latexsym,calc,amssymb,amsfonts}
\usepackage{amsmath, enumerate, algorithm, algorithmic}
\usepackage{graphics,graphicx}

 \newtheorem{thm}{Theorem}
 \newtheorem{lemma}[thm]{Lemma}
 
 \newtheorem{cor}[thm]{Corollary}
\newtheorem{dfn}{Definition}
\newtheorem{remark}{Remark}

\newproof{pf}{Proof}

\def\a{\ensuremath{\alpha}}
\def\b{\ensuremath{\beta}}
\def\g{\ensuremath{\gamma}}
\def\d{\ensuremath{\delta}}
\def\r{\ensuremath{\rho}}
\def\e{\ensuremath{\epsilon}}

\def\w{\ensuremath{\omega}}

\def\RR{\mathbb{R}}

\def\O{\emptyset}

\def\kn{\frac{k}{n}}
\def\nN{\frac{n}{N}}

\def\lone{\ell_1}

\def\xTTL{x_{T-T^l}}
\def\xTTLL{x_{T-T^{l-1}}}

\def\msp{\mu^{sp}(k,n,N)}
\def\mcsp{\mu^{csp}(k,n,N)}
\def\miht{\mu^{iht}(k,n,N)}
\def\mromp{\mu^{r}(k,n,N)}

\def\mspdr{\mu^{sp}(\d,\r)}
\def\mcspdr{\mu^{csp}(\d,\r)}
\def\mihtdr{\mu^{iht}(\d,\r)}

\def\xispk{\xi^{sp}(k,n,N)}

\def\xispdr{\xi^{sp}(\d,\r)}

\def\xicsp{\xi^{csp}(k,n,N)}

\def\xicspdr{\xi^{csp}(\d,\r)}

\def\xiiht{\xi^{iht}(k,n,N)}
\def\xiihtk{\xi^{iht}(k,n,N)}
\def\xiihtdr{\xi^{iht}(\d,\r)}

\def\rfl1{\rho_S^{\lone}(\delta)}

\def\rflqc1{\rho_S^{\lone}(\delta;q,C_1(\delta,\rho)\le\Upsilon)}

\def\rcsp{\rho_S^{csp}(\delta)}
\def\riht{\rho_S^{iht}(\delta)}
\def\rsp{\rho_S^{sp}(\delta)}

\def\lmax{\lambda_{max}}
\def\lmin{\lambda_{min}}

\def\nux{\nu_{\text{min}}(x)}

\newcommand{\argmin}{\operatornamewithlimits{arg\,min}}

\begin{document}

\title{Phase Transitions for Greedy Sparse Approximation Algorithms}

%
%

\author[gr]{Jeffrey D. Blanchard\corref{cor1}\fnref{fn1}}
\ead{jeff@math.grinnell.edu}
\author[ed]{Coralia Cartis}
\ead{coralia.cartis@ed.ac.uk}
\author[ed]{Jared Tanner\fnref{fn2}}
\ead{jared.tanner@ed.ac.uk}
\author[ed]{Andrew Thompson}
\ead{a.thompson-8@sms.ed.ac.uk}

\cortext[cor1]{Corresponding author}
\fntext[fn1]{JDB was supported by NSF DMS grant 0602219 while
  a VIGRE postdoctoral fellow at Department of Mathematics, University of Utah.}
\fntext[fn2]{JT acknowledges support from the Philip Leverhulme and
  the Alfred P. Sloan Fellowships.}

\address[gr]{Department of Mathematics and Statistics, Grinnell College, Grinnell, Iowa 50112-1690, USA.}
\address[ed]{School of Mathematics and the Maxwell Institute, University of Edinburgh, King's Buildings, Mayfield Road, Edinburgh EH9 3JL, UK.}

\begin{abstract}
A major enterprise in compressed sensing 
and sparse approximation is the design and analysis of computationally
tractable algorithms for recovering sparse, exact or approximate, solutions of
underdetermined linear systems of equations.  Many such algorithms have now
been proven to have optimal-order uniform recovery guarantees using the ubiquitous 
Restricted Isometry Property (RIP) \cite{CTdecoding}. However, it is
 unclear when the RIP-based sufficient conditions on the algorithm are
satisfied. We present a framework in which this task can be
achieved; translating these conditions for Gaussian measurement
matrices into requirements on the signal's sparsity level, length, and number of
measurements. We illustrate this approach on three of the state-of-the-art  greedy algorithms:
CoSaMP \cite{NeTr09_cosamp}, Subspace Pursuit (SP)
\cite{SubspacePursuit} and Iterative Hard Thresholding (IHT)
\cite{BlDa08_iht}. Designed to allow a direct comparison of existing theory, 
our framework implies that, according to the best known bounds, IHT requires the fewest 
number of compressed sensing measurements and has the lowest per iteration computational 
cost of the three algorithms compared here.
\end{abstract}


\maketitle

\thispagestyle{plain}
\markboth{J. D. BLANCHARD, C. CARTIS, J. TANNER, and
  A. THOMPSON}{Phase Transitions for Greedy Sparse Approximation Algorithms}

\section{Introduction}\label{sec:intro}

In compressed sensing
\cite{CompressiveSampling,CTdecoding,CompressedSensing}, one works
under the sparse approximation assumption, namely, that signals/vectors of interest can be well approximated by few components
of a known basis.  This assumption is often satisfied due to
constraints imposed by the system which generates the signal.  In this
setting, it has been proven (originally in \cite{CTdecoding,
  CompressedSensing} and by many others since) that the number of
linear observations of the signal, required to guarantee recovery, 
 need only be proportional to the sparsity of the signal's approximation.  This is in stark contrast to the standard Shannon-Nyquist Sampling paradigm \cite{shannon} where worst-case sampling requirements are imposed.  

In the simplest setting, consider measuring a vector $x_0\in\RR^N$
which either has exactly $k<N$ nonzero entries, or which has $k$
entries whose magnitudes are dominant.  Let $A$ be an $n\times N$
matrix with $n<N$ which we use to measure $x_0$; the $n$ inner
products with $x_0$ are the entries in $y=Ax_0$.  From knowledge of
$y$ and $A$ one seeks to recover the vector $x_0$, or a suitable
approximation thereof, \cite{CS_SIREV}.  Let $\chi^N(k):=\left\{x\in\RR^n:
  \left\|x\right\|_0\le k\right\}$ denote the family of at most $k$-sparse vectors in $\RR^N$, where $\left\|\cdot\right\|_0$ counts the number of nonzero entries.  From $y$ and $A$, the optimal $k$-sparse signal is the solution of
\begin{equation}\label{eq:l0}
\min_{x\in\chi^N(k)}\left\|Ax-y\right\|_2,
\end{equation}
where $\|\cdot\|_2$ denotes the Euclidean norm.

However, solving \eqref{eq:l0} via a naive exhaustive search is
combinatorial in nature and NP-hard \cite{NPhard}.  A major aspect of compressed sensing 
theory is the study of alternative methods to solving
\eqref{eq:l0}.  Since the system $y=Ax$ is underdetermined, any
successful recovery of $x$ will require some form of nonlinear
reconstruction.  Under certain conditions, various algorithms have been shown to successfully reduce \eqref{eq:l0} to a tractable problem, one with a computational cost which is a low degree 
polynomial of the problem dimensions, rather than the exponential cost associated with a direct combinatorial search for the solution of \eqref{eq:l0}.  While there are numerous reconstruction algorithms, they each generally fall into one of three categories: \emph{greedy methods}, \emph{regularizations}, or \emph{combinatorial group testing}. For an indepth discussion of compressed sensing recovery algorithms, see \cite{NeTr09_cosamp} and references therein. 

The first uniform guarantees for exact reconstruction of every
$x\in\chi^N(k)$, for a fixed $A$, came from $\lone$-regularization.  In this case, \eqref{eq:l0} is relaxed to solving the problem
\begin{equation}\label{eq:l1}
\min_{x\in\RR^N}\left\|x\right\|_1 \ \hbox{subject to}\ \left\|Ax-y\right\|_2<\g,
\end{equation}
for some known noise level, or decreasing, $\g$.
$\lone$-regularization has been extensively studied, see the 
pioneering works \cite{CTdecoding, CompressedSensing}; also, see
\cite{Do05_signal,DoTa08_JAMS,RIconstants} for results analogous to
those presented here.  In this paper, we focus on three illustrative greedy
algorithms, \emph{Compressed Sensing Matching Pursuit} (CoSaMP)
\cite{NeTr09_cosamp}, \emph{Subspace Pursuit} (SP) \cite{SubspacePursuit},
and \emph{Iterative Hard Thresholding} (IHT)
\cite{BlDa08_iht}, which boast similar uniform guarantees of
successful recovery of sparse signals when the measurement matrix $A$
satisfies the now ubiquitous \emph{Restricted Isometry Property} (RIP)
\cite{CTdecoding, RIconstants}. The three algorithms are deeply connected and each
have some advantage over the other.  These algorithms are essentially
support set recovery algorithms which use hard thresholding to
iteratively update the approximate support set; their differences lie
in the magnitude of the application of hard thresholding and the
vectors to which the thresholding is applied, \cite{DoMa09,TrWr09}.
The algorithms are restated in the next section. Other greedy methods
with similar guarantees are available, see for example \cite{RON, IHTother};
several other greedy techniques have been developed (\cite{DTDS08, NeVe06_UUP,OMP_wakin}, etc.), but their theoretical analyses do not currently subscribe to the above
uniform framework.

As briefly mentioned earlier, the intriguing aspect of compressed sensing
is its ability to recover $k$-sparse signals when the number of
measurements required is proportional to the sparsity, $n\sim k$, as
the problem size grows, $n\rightarrow\infty$.  Each of the algorithms
discussed here exhibit a phase transition property, where there exists
a $k_n^{\ast}$ such that for any $\epsilon>0$, as
$k_n^{\ast},n\rightarrow\infty$, the algorithm successfully recovers
all $k$-sparse vectors provided $k<(1-\epsilon)k^{\ast}_n$ and does
not recover all $k$-sparse vectors if $k>(1+\epsilon)k_n^{\ast}$.
For a description of phase transitions in the context of compressed
sensing, see \cite{PUT}, while for numerical average-case phase
transitions for greedy algorithms, see  \cite{DoMa09}.
We consider the asymptotic setting where $k$ and $N$ grow proportionally
with $n$, namely, $(k,n,N)\rightarrow\infty$ with the ratios
$\frac{k}{n}\rightarrow\r, \frac{n}{N}\rightarrow\d$ as $n\rightarrow\infty$ for 
$(\delta,\rho)\in (0,1)^2$; also, we assume the matrix $A$ is
drawn i.i.d.\ from $\mathcal{N}(0,n^{-1})$, the normal distribution
with mean $0$ and variance $n^{-1}$. In this framework, we develop
lower bounds on the phase transition for exact recovery of all $k$-sparse
signals. These bounds provide curves in the unit square, $(\delta,\rho)\in (0,1)^2$, below which
there is an exponentially high probability on the draw the Gaussian
matrix $A$, that $A$ will satisfy the sufficient RIP conditions and
therefore solve \eqref{eq:l0}.  We utilize a more general, asymmetric
version of the RIP, see Definition \ref{def:LU}, to compute as precise
a lower bound on the phase transitions as possible.  This phase
transition framework allows a direct comparison of the provable
recovery regions of different algorithms in terms of the problem instance $(\nN,\kn)$ .  We then compare the guaranteed recovery capabilities of these algorithms to the guarantees of $\lone$-regularization proven via RIP analysis. For $\lone$-regularization, this phase transition framework has already been applied using the RIP \cite{BlCaTa09_spars,RIconstants}, using the theory of convex polytopes \cite{Do05_signal} and geometric functional analysis \cite{RV07_gaussian}.

\begin{figure}[t]
 
\begin{center}
\begin{tabular}{cc}
 \includegraphics[bb= 70 215 546 589, width=2.70 in,height=2.00 in]{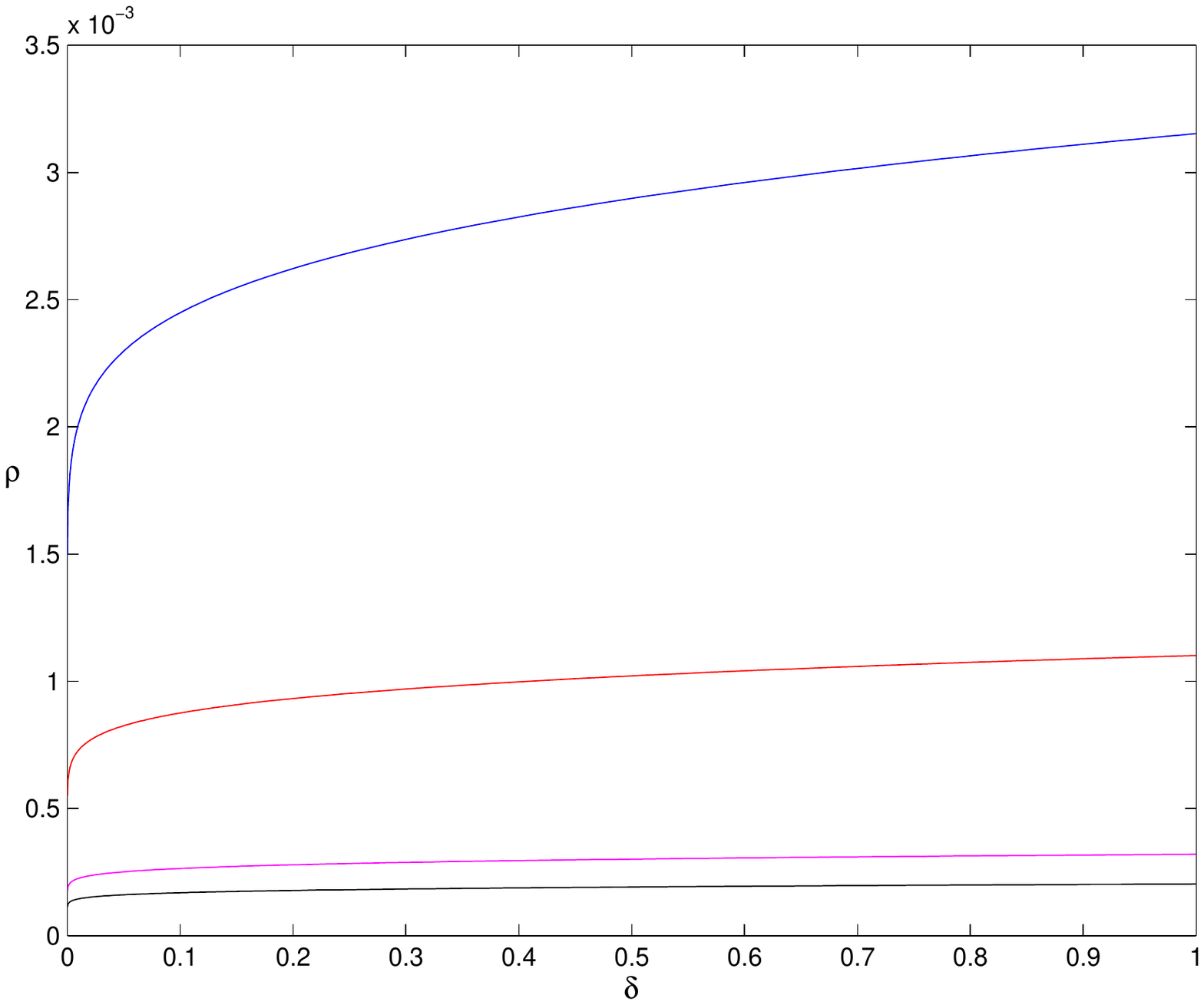}
&
\includegraphics[bb= 70 215 546 589, width=2.70 in,height=2.00 in]{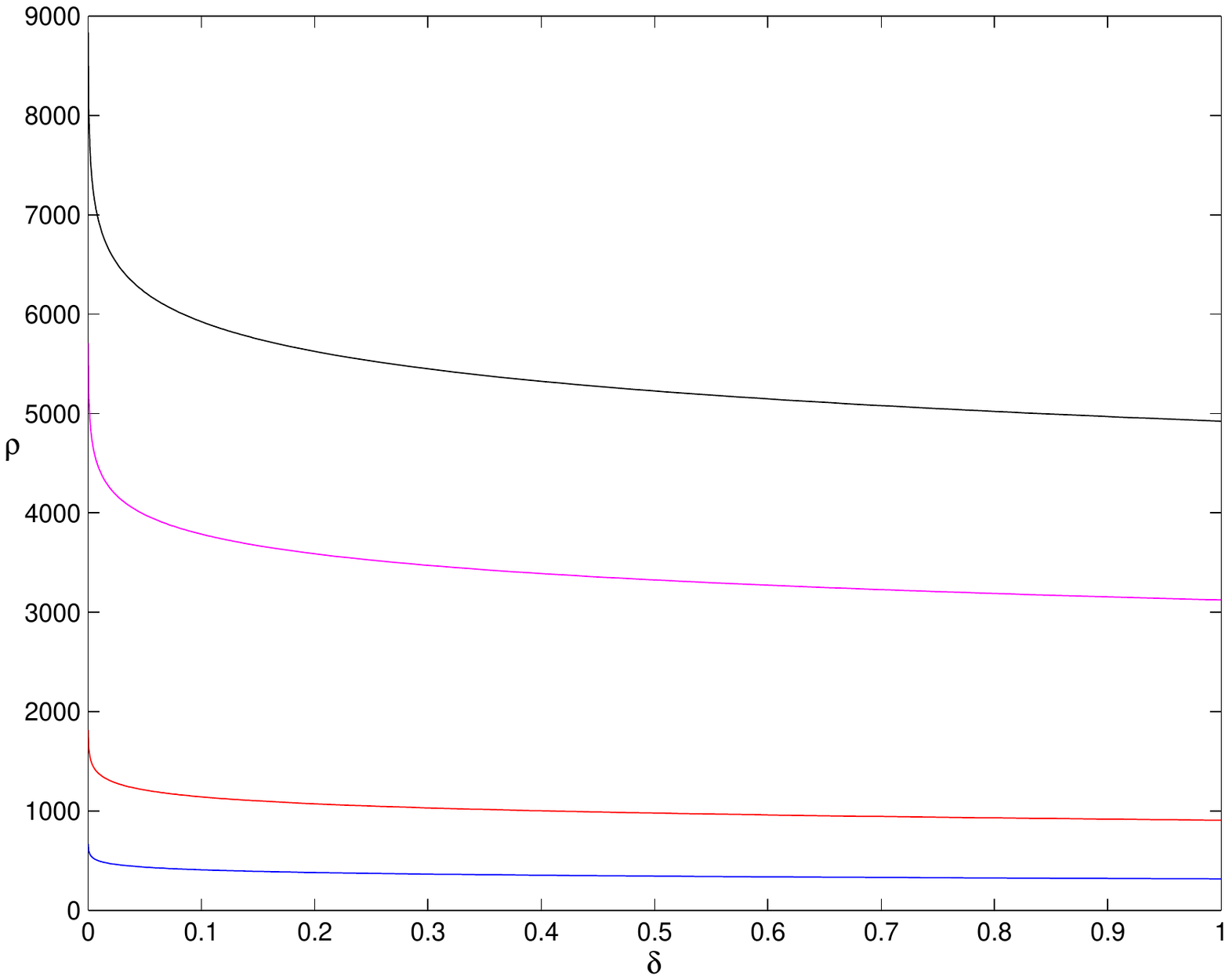}\\
(a) & (b)\\
\end{tabular}
 \caption{(a): The lower bounds on the Strong  exact recovery phase
   transition for Gaussian random matrices for the algorithms
   $\lone$-regularization (Theorem \ref{thm:Lonenoisyphase}, $\rfl1$, blue), IHT (Theorem \ref{thm:IHTnoisyphase}, $\riht$, red),
   SP (Theorem \ref{thm:SPnoisyphase}, $\rsp$, magenta),
   and CoSaMP (Theorem \ref{thm:CSPnoisyphase}, $\rcsp$, black).  (b): The inverse of the phase transition lower bounds in
 the left panel (a).\label{fig:transitions}}
 \end{center}
 \end{figure}

The aforementioned lower bounds on the algorithmic exact sparse recovery phase transitions are presented in Theorems \ref{thm:CSPnoisyphase}, \ref{thm:SPnoisyphase}, and \ref{thm:IHTnoisyphase}.  The curves are defined by functions $\rsp$ (SP; the magenta curve in Fig.\ref{fig:transitions}(a)), $\rcsp$ (CoSaMP; the black curve in Fig.\ref{fig:transitions}(a)), $\riht$ (IHT; the red curve in Fig.\ref{fig:transitions}(a)).  For comparison, the analogous lower bound on the phase transition for $\rfl1$ ($\lone$-regularization) is displayed as the blue curve in Fig.\ref{fig:transitions}(a).  From Fig.~\ref{fig:transitions}, we are able to directly compare the provable recovery results of the three greedy algorithms as well as $\lone$-regularization.  For a given problem instance $(k,n,N)$ with the entries of $A$ drawn i.i.d. from $\mathcal{N}(0,n^{-1})$, if $\kn=\r$ falls in the region below the curve $\r_S^{alg}(\d)$ associated to a specific algorithm, then with probability approaching 1 exponentially in $n$, the algorithm will exactly recover the $k$-sparse vector $x\in\chi^N(k)$ no matter which $x\in\chi^N(k)$ was measured by $A$.
These lower bounds on the phase transition can also be interpreted as
the minimum number of measurements known to guarantee recovery through
the constant of proportionality: $n>\left(\r_S^{alg}\right)^{-1} k$.
Fig.~\ref{fig:transitions}(b) portrays the inverse of the lower bounds
on the phase transition.  This gives a minimum possible value for
$\left(\r_S^{alg}\right)^{-1}$.  For example, from the blue curve, for
a Gaussian random matrix used in $\lone$-regularization, the minimum
number of measurements proven (using RIP) to be sufficient to ensure recovery
of all  $k$-sparse vectors is $n > 317 k$.  By contrast, for  greedy algorithms, the minimum number of measurements shown to be sufficient is significantly larger: $n>907k$ for IHT, $n>3124k$ for SP, and $n>4923k$ for CoSaMP.

More precisely, the main contributions of this article is the derivation 
of theorems and corollaries of the following form for each of the
CoSaMP, SP, and IHT algorithms.

\begin{thm}\label{thm:alg}
Given a matrix $A$ with entries drawn i.i.d. from ${\cal N}(0,n^{-1})$, for
any  $x\in\chi^N(k)$, let $y=Ax+e$ for some (unknown) noise vector $e$.
For any $\epsilon\in (0,1)$, as $(k,n,N)\rightarrow\infty$ with
$n/N\rightarrow\d\in(0,1)$ and $k/n\rightarrow\r\in (0,1)$, there
exists $\mu^{alg}(\d,\r)$ and $\r_S^{alg}(\d)$, the unique solution to
$\mu^{alg}(\d,\r)=1$. If  $\r<(1-\e)\r_S^{alg}(\d)$, 
there is an exponentially high probability on the draw
of $A$ that the output of the algorithm at the $l^{th}$ iteration, $\hat{x}$, approximates $x$ within the  bound
\begin{equation}\label{eq:SPnoisyphase}
\|x-\hat{x}\|_2 \le \kappa^{alg}(\d,(1+\e)\r)\left[\mu^{alg}(\d,(1+\e)\r)\right]^l \|x\|_2 + \frac{\xi^{alg}(\d,(1+\e)\r)}{1-\mu^{alg}(\d,(1+\e)\r)}\|e\|_2,
\end{equation}  
for some $\kappa^{alg}(\d,\r)$ and $\xi^{alg}(\d,\r)$.
\end{thm}

\begin{cor}\label{cor:alg}
Given a matrix $A$ with entries drawn i.i.d. from ${\cal N}(0,n^{-1})$, for
any  $x\in\chi^N(k)$, let $y=Ax$.
For any $\epsilon\in (0,1)$, with $n/N\rightarrow\d\in(0,1)$ and
$k/n\rightarrow\r<(1-\e)\r_S^{alg}(\d)$ as $(k,n,N)\rightarrow\infty$, there
is an exponentially high probability on the draw of $A$ that the algorithm 
exactly recovers $x$ from $y$ and $A$ in a finite number of
iterations not to exceed 
\begin{equation}
\ell^{alg}_{max}(x):=\left\lceil \frac{\log\nux-\log\kappa^{alg}(\d,\r)}{\log\mu^{alg}(\d,\r)}\right\rceil+1
\end{equation}
where
\begin{equation}\label{eq:numin}
\nux:= \frac{\min_{i\in T} |x_i|}{\left\|x\right\|_2}
\end{equation}
with $T:=\{i: x_i\ne 0\}$ and $\lceil m\rceil$, the smallest integer greater than or equal to $m$.
\end{cor}

\begin{figure}[p]
\begin{center}
\begin{tabular}{cc}
\includegraphics[bb= 70 215 546 589, width=2.70 in,height=2.00 in]{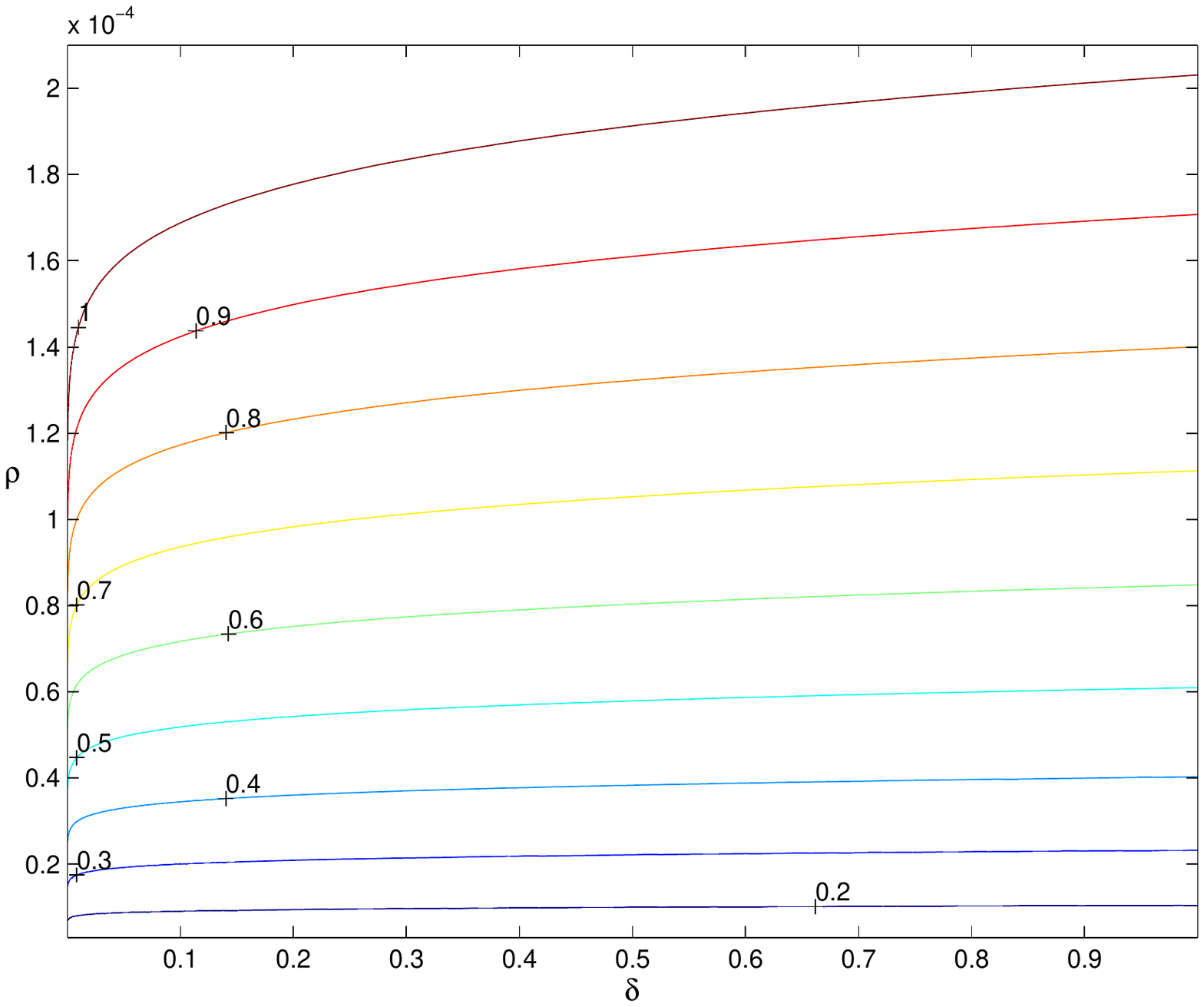} &
\includegraphics[bb= 70 215 566 589, width=2.70 in,height=2.00 in]{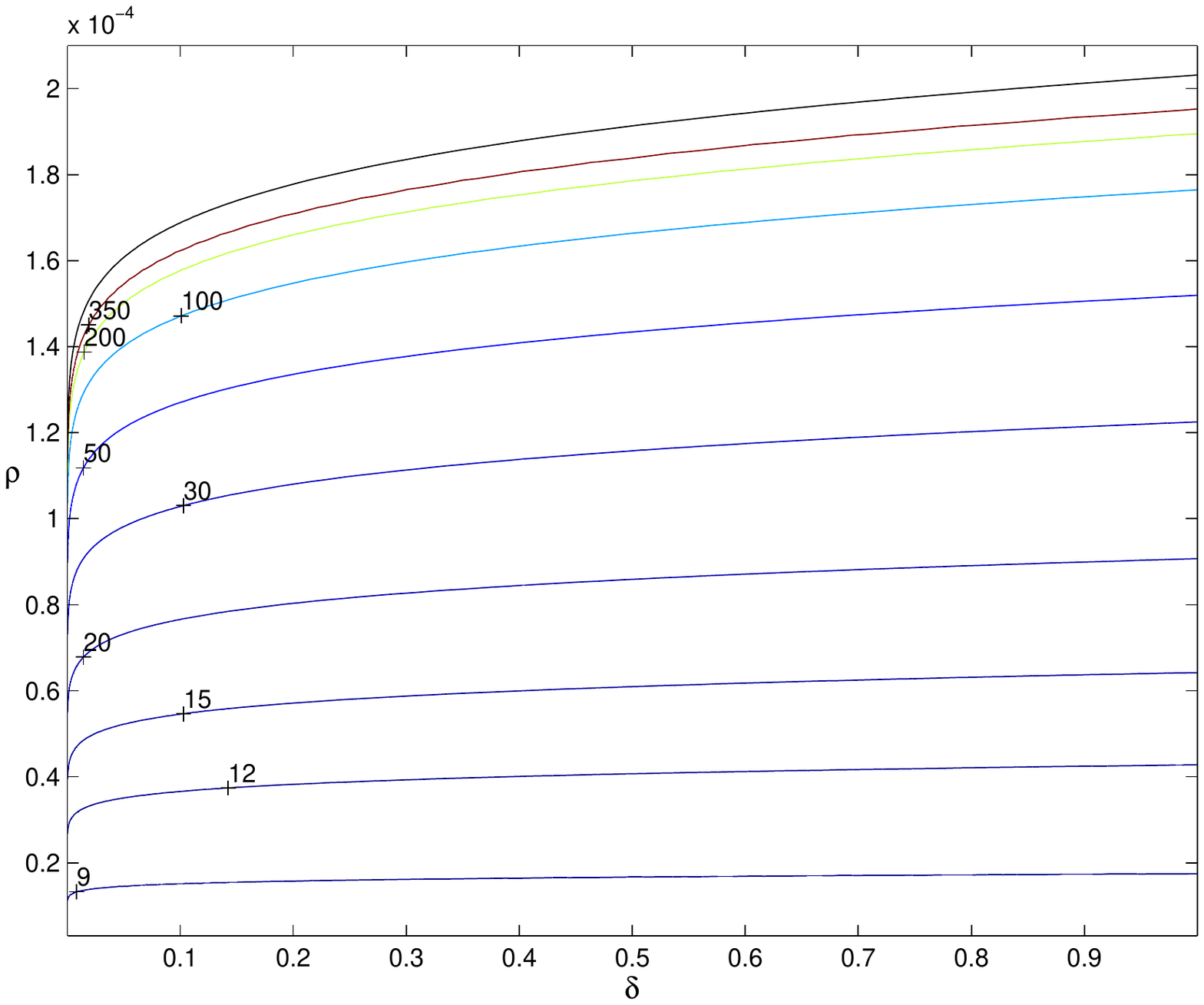}  \\
(a) & (b) \\
 & \\
\includegraphics[bb= 70 215 546 589, width=2.70 in,height=2.00 in]{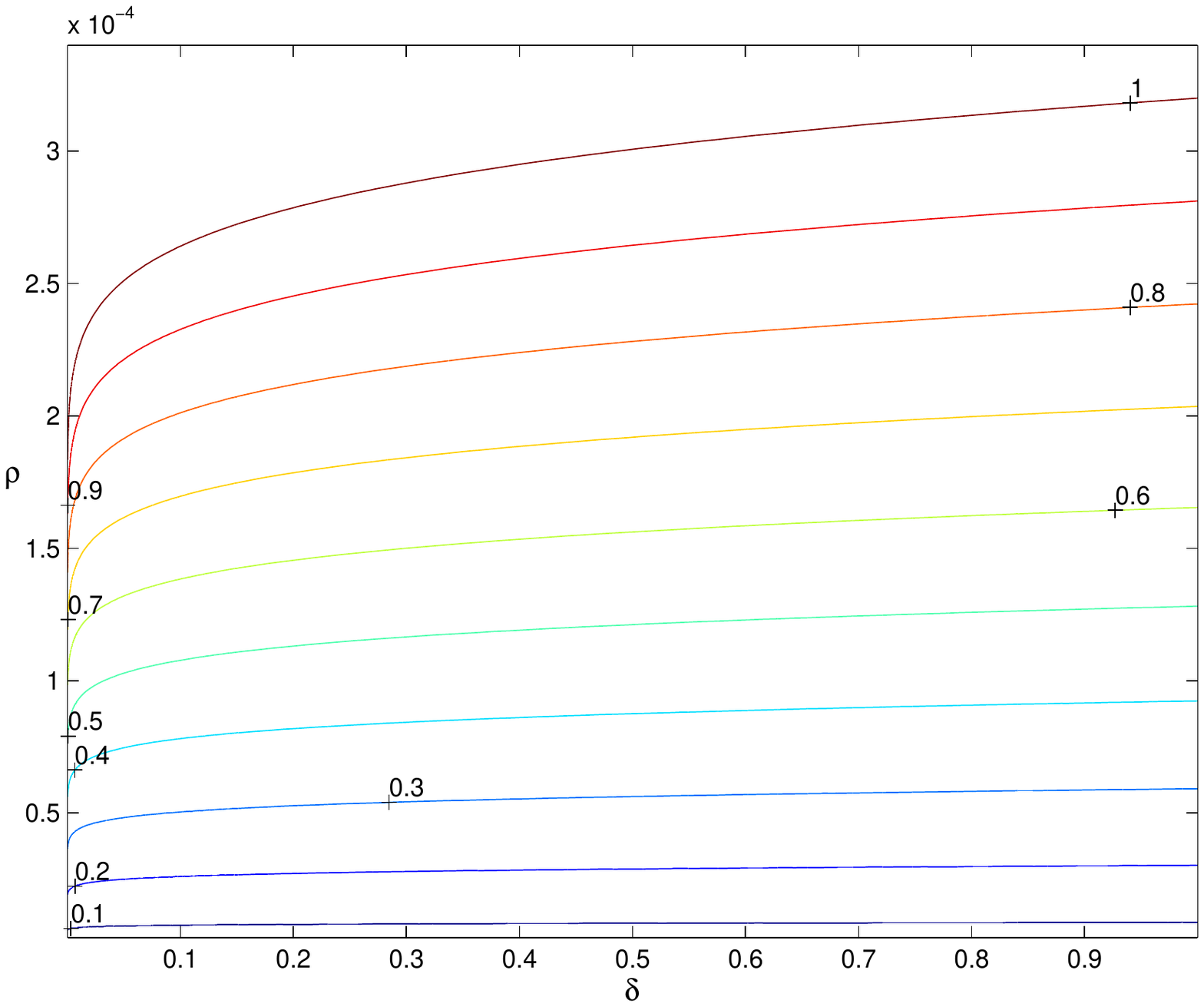} &
\includegraphics[bb= 70 215 566 589, width=2.70 in,height=2.00 in]{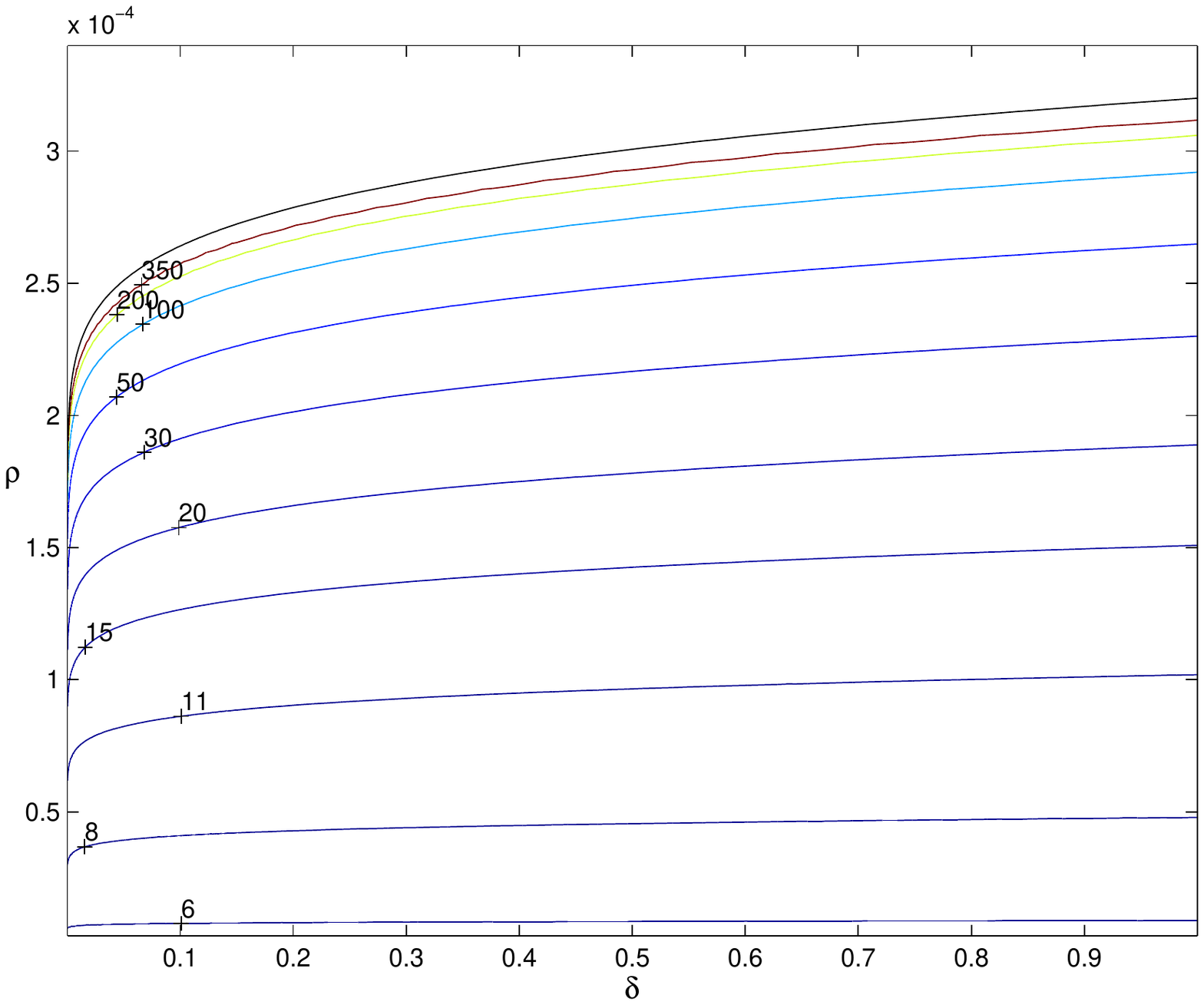} \\
(c) & (d) \\
 & \\
\includegraphics[bb= 70 215 546 589, width=2.70 in,height=2.00 in]{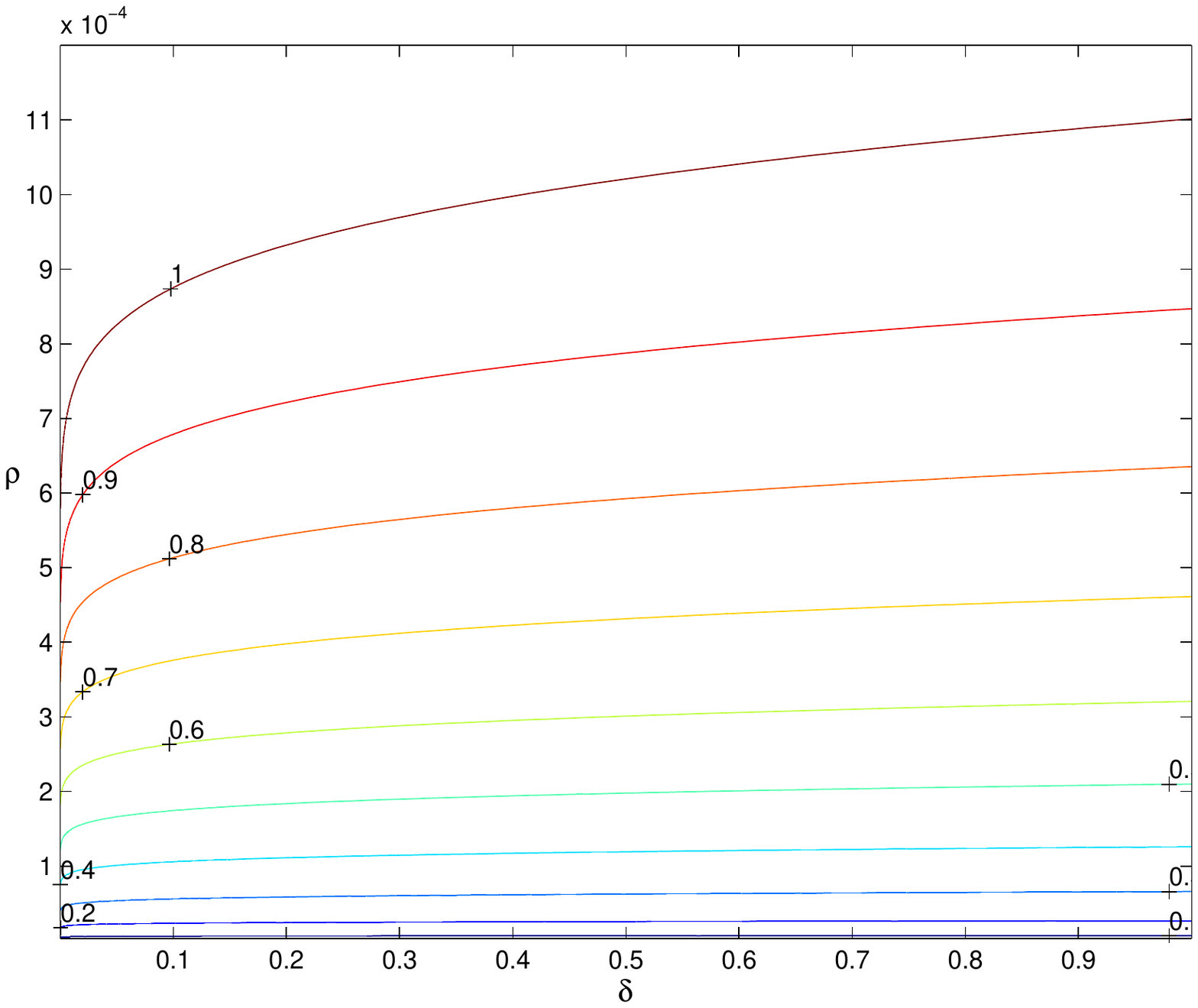} &
\includegraphics[bb= 70 215 566 589, width=2.70 in,height=2.00 in]{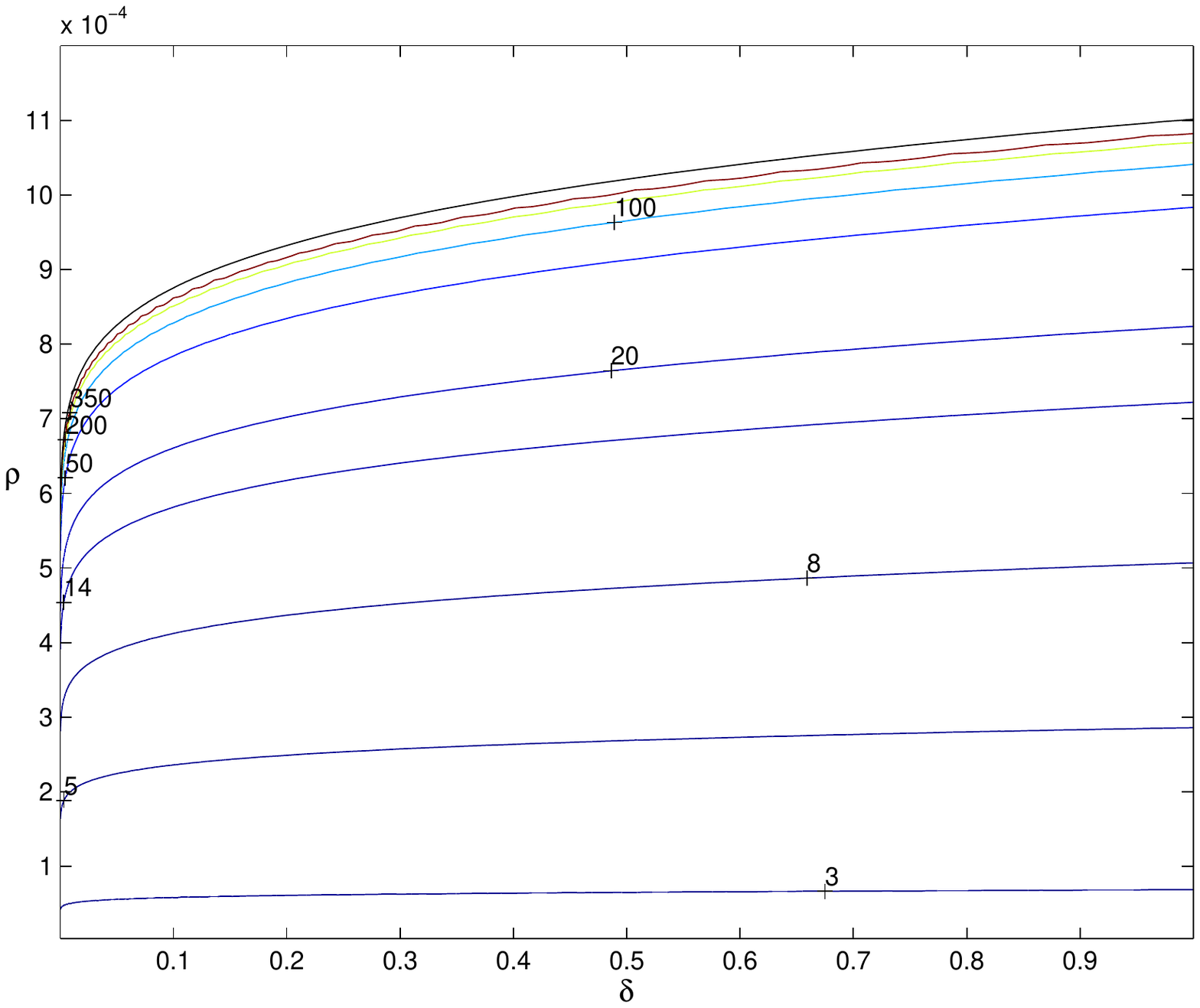} \\
(e) & (f) \\
\end{tabular}
 \caption{For CoSaMP (a-b), SP (c-d), and IHT (e-f) from the left to the right panel: the convergence factor $\mu^{alg}(\d,\r)$ and the stability factor $\frac{\xi^{alg}}{1-\mu^{alg}}(\d,\r)$.}\label{fig:Stab_sp_csp}
 \end{center}
 \end{figure}

The factors $\mu^{alg}(\d,\r)$ and $\frac{\xi^{alg}}{1-\mu^{alg}}(\d,\r)$ for CoSaMP, SP, and IHT are
displayed in Figure \ref{fig:Stab_sp_csp}, while formulae for their calculation are deferred to Section \ref{sec:PhaseTrans}.

Corollary \ref{cor:alg} implies that
$\r^{alg}_S(\d)$ delineates the region in which the algorithm can be
guaranteed to converge provided  there exists an $x\in\chi^N(k)$ such
that $y=Ax$. 
However, if no such $x$ exists, as $\r$ approaches $\rho^{alg}_S(\d)$ the guarantees on the number of iteraties required and stability factors become unbounded.
Further bounds on the convergence factor $\mu^{alg}(\d,\r)$ and the
stability factor $\frac{\xi^{alg}}{1-\mu^{alg}}(\d,\r)$ result in yet
lower curves $\rho_S^{alg}(\d;bound)$ for a specified $bound$; recall
that $\rho^{alg}_S(\d)$ corresponds to the bound $\mu^{alg}(\d,\r)=1$.

In the next section, we recall the three algorithms and introduce
necessary notation.  Then we present the asymmetric restricted
isometry property and formulate weaker restricted isometry conditions
on a matrix $A$ that ensure the respective algorithm will successfully
recover all $k$-sparse signals. In addition to exact recovery, we
study the known bounds on the behavior of the algorithms in the presence 
of noisy measurements. 
In order to make quantitative comparisons of these results, we
must select a matrix ensemble for analysis.  In Section
\ref{sec:PhaseTrans}, we present the lower bounds on the phase
transition for each algorithm when the measurement matrix is a
Gaussian random matrix.  Phase transitions are developed in the case
of exact sparse signals while bounds on the multiplicative stability
constants are also compared through associated level curves.  Section
\ref{sec:discussion} is a discussion of our interpretation of these
results and how to use this phase transition framework for comparison
of other  algorithms.  

For an index set $I\subset\{1,\dots,N\}$, let $x_I$ denote the restriction of a vector $x\in\RR^N$ to the set $I$, i.e., $(x_I)_i=x_i$ for $i\in I$ and $(x_I)_j=0$ for $j\notin I$.  Also, let $A_I$ denote the submatrix of $A$ obtained by selecting the columns $A$ indexed by $I$.  $A_I^*$ is the conjugate transpose of $A_I$ while $A_I^\dagger=(A_I^*A_I)^{-1}A_I^*$ is the pseudoinverse of $A_I$.  In each of the algorithms, thresholding is applied by selecting $m$ entries of a vector with largest magnitude; we refer to this as hard thresholding of magnitude $m$.

\section{Greedy Algorithms and the Asymmetric Restricted Isometry Property}\label{sec:ARIP}

\subsection{CoSaMP}\label{sec:algCSP}

The CoSaMP recovery algorithm is a support recovery algorithm which
applies hard thresholding by selecting the $k$ largest entries of a
vector obtained by applying a pseudoinverse to the measurement $y$.
In CoSaMP, the columns of $A$ selected for the pseudoinverse are
obtained by applying hard thresholding of magnitude $2k$ to $A^*$
applied to the residual from the previous iteration and adding these
indices to the approximate support set from the previous iteration.
This larger pseudoinverse matrix of size $2k\times n$ imposes the most
stringent aRIP condition of the three algorithms.  However, CoSaMP
uses one fewer pseudoinverse per iteration than SP as the residual vector is computed with a direct matrix-vector multiply of size $n\times k$ rather than with an additional pseudoinverse. Furthermore, when computing the output vector $\hat{x}$, CoSaMP does not need to apply another pseudoinverse as does SP.  See Algorithm~\ref{alg:CSP}.  

\begin{algorithm}[h!]
\caption{CoSaMP \cite{NeTr09_cosamp}}\label{alg:CSP}
\textbf{Input:} $A$, $y$, $k$

\underline{\textbf{Output:} A $k$-sparse approximation $\hat{x}$ of the target signal $x$ }
 
\medskip
\textbf{Initialization:}

\begin{algorithmic}[1]
\STATE Set $T^0=\O$
\STATE Set $y^0=y$
\end{algorithmic}

\medskip
\textbf{Iteration:} During iteration $l$, \textbf{do}

\begin{algorithmic}[1]
\STATE $\tilde{T}^l= T^{l-1}\cup\{2k\ \hbox{indices of largest magnitude entries of}\ A^*y^{l-1}\}$
\STATE $\tilde{x}=A_{\tilde{T}^l}^\dagger y$
\STATE $T^l=\{k\ \hbox{indices of largest magnitude entries of}\ \tilde{x}\}$
\STATE $y^l=y-A_{T^l}\tilde{x}_{T^l}$
\IF {$\|y^l\|_2 = 0$}
\RETURN $\hat{x}$ defined by $\hat{x}_{\{1,\dots,N\}-T^l}=0$ and $\hat{x}_{T^l}=\tilde{x}_{T^l}$
\ELSE
\STATE Perform iteration $l+1$
\ENDIF
\end{algorithmic}
\end{algorithm}

\subsection{Subspace Pursuit}\label{sec:algSP}

The Subspace Pursuit algorithm is also a support recovery algorithm  which applies hard thresholding of magnitude $k$ to a vector obtained by applying a pseudoinverse to the measurements $y$.   The submatrix chosen for the pseudoinverse has its columns selected by applying $A^*$ to the residual vector from the previous iteration, hard thresholding of magnitude $k$, and adding the indices of the terms to the previous approximate support set.  Compared to the other two algorithms, a computational disadvantage of SP is that the aforementioned residual vector is also computed via a pseudoinverse, this time selecting the columns from $A$ by again applying a hard threshold of magnitude $k$. The computation of the approximation to the target signal also requires the application of a pseudoinverse for a matrix of size $n\times k$. See Algorithm~\ref{alg:SP}.  

\algsetup{indent=2em}

\begin{algorithm}[h!]
\caption{Subspace Pursuit \cite{SubspacePursuit}}\label{alg:SP}
\textbf{Input:} $A$, $y$, $k$

\underline{\textbf{Output:} A $k$-sparse approximation $\hat{x}$ of the target signal $x$ }
 
\medskip
\textbf{Initialization:}

\begin{algorithmic}[1]
\STATE Set $T^0=\{k\ \hbox{indices of largest magnitude entries of}\ A^*y\}$
\STATE Set $y_r^0=y-A_{T^0}A_{T^0}^\dagger y$
\end{algorithmic}

\medskip
\textbf{Iteration:} During iteration $l$, \textbf{do}

\begin{algorithmic}[1]
\STATE $\tilde{T}^l=T^{l-1}\cup \{k\ \hbox{indices of largest magnitude entries of}\ A^*y_r^{l-1}\}$
\STATE Set $\tilde{x}=A_{\tilde{T}^l}^\dagger y$
\STATE $T^l=\{k\ \hbox{indices of largest magnitude entries of}\ \tilde{x}\}$
\STATE $y_r^l=y-A_{T^l}A_{T^l}^\dagger y$
\IF {$\|y_r^l\|_2 = 0$}
\RETURN $\hat{x}$ defined by $\hat{x}_{\{1,\dots,N\}-T^l}=0$ and $\hat{x}_{T^l}=A_{T^l}^\dagger y$
\ELSE
\STATE Perform iteration $l+1$
\ENDIF
\end{algorithmic}
\end{algorithm}

\subsection{Iterative Hard Thresholding}\label{sec:algIHT}

Iterative Hard Thresholding (IHT) is also a support recovery
algorithm.  However, IHT applies hard thresholding to an approximation
of the target signal, rather than to the residuals.  This completely
eliminates the use of a pseudoinverse, reducing the computational
cost per iteration.  In particular, hard thresholding of magnitude $k$ is applied to
an updated approximation of the target signal, $x$, obtained by
matrix-vector multiplies of size $n\times N$ that represent a move by
a fixed stepsize $\omega$ along the steepest
descent direction from the current iterate for the residual $\|Ax-y\|_2^2$.
See Algorthm \ref{alg:IHT}.

\begin{algorithm}[h!]
\caption{Iterative Hard Thresholding \cite{BlDa08_iht}}\label{alg:IHT}
\textbf{Input:} $A$, $y$, $\omega \in (0,1)$, $k$

\underline{\textbf{Output:} A $k$-sparse approximation $\hat{x}$ of the target signal $x$ }
 
\medskip
\textbf{Initialization:}

\begin{algorithmic}[1]
\STATE Set $x^0=0$
\STATE Set $T^0=\O$
\STATE Set $y^0=y$
\end{algorithmic}

\medskip
\textbf{Iteration:} During iteration $l$, \textbf{do}

\begin{algorithmic}[1]

\STATE $x^l= x^{l-1}_{T^{l-1}}+ w A^*y^{l-1}$
\STATE $T^l=\{k\ \hbox{indices of largest magnitude entries of}\ x^l\}$
\STATE $y^l=y-A_{T^l}x_{T^l}^l$
\IF {$\|y^l\|_2 = 0$}
\RETURN $\hat{x}$ defined by $\hat{x}_{\{1,\dots,N\}-T^l}=0$ and $\hat{x}_{T^l}=x_{T^l}^l$ 
\ELSE
\STATE Perform iteration $l+1$
\ENDIF
\end{algorithmic}
\end{algorithm}

\begin{remark}
({\bf Stopping criteria for greedy methods})\quad
In the case of corrupted measurements, where $y=Ax+e$ for some noise
vector $e$, the stopping criteria listed in Algorithms
\ref{alg:CSP}-\ref{alg:IHT} may never be achieved.  Therefore, a
suitable alternative stopping criteria must be employed.  For our
analysis on bounding the error of approximation in the noisy case, we
bound the approximation error if the algorithm terminates after $l$
iterations.  For example, we could change the algorithm to require a
maximum number of iterations $l$ as an input and then terminate the
algorithm if our stopping criteria is not met in fewer iterations.  In
practice, the user would be better served to stop the algorithm when
the residual is no longer improving.  For a more thorough discussion
of suitable stopping criteria for each algorithm in the noisy case,
see the original announcement of the algorithms
\cite{BlDa08_iht,SubspacePursuit,NeTr09_cosamp}.
\end{remark}

\subsection{The Asymmetric Restricted Isometry Property}

In this section we relax the sufficient conditions originally placed on Algorithms \ref{alg:CSP}-\ref{alg:IHT} by employing a more general notion of a restricted isometry.  As discussed in \cite{RIconstants}, the singular values of the $n\times k$ submatrices of an arbitrary measurement matrix $A$ do not, in general, deviate from unity symmetrically.  The standard notion of the \emph{restricted isometry property} (RIP) \cite{CTdecoding} has an inherent symmetry which is unneccessarily restrictive. Hence, seeking the best possible conditions for the measurement matrix under which Algorithms \ref{alg:CSP}-\ref{alg:IHT} will provably recovery every $k$ sparse vector, we reformulate the sufficient conditions in terms of the \emph{asymmetric restricted isometry property} (aRIP) \cite{RIconstants}.

\begin{dfn}\label{def:LU}
For an $n\times N$ matrix $A$, the \emph{asymmetric RIP constants} 
$L(k,n,N)$ and $U(k,n,N)$ are defined as: 
\begin{align}
L(k,n,N)&:=\min_{c\ge0}\,c\,\,\,\mbox{subject to}\,\,\,
 (1-c)\|x\|_2^2 \le \|Ax\|_2^2, \,\,\forall x\in\chi^N(k);\label{eq:Lrip} \\
U(k,n,N)&:=\min_{c\ge0} \,c\,\,\,\mbox{subject to}\,\,\,
(1+c) \|x\|_2^2 \ge \left\|Ax\right\|_2^2, \,\,\forall x\in\chi^N(k).\label{eq:Urip}
\end{align}
\end{dfn}

\begin{remark}\label{rem:ripnondecreasing}
\begin{enumerate}
\item
The more common, symmetric definition of the RIP constants is recovered by defining $R(k,n,N)=\max\{L(k,n,N),U(k,n,N)\}$.  In this case, a matrix $A$ of size $n\times N$ has the RIP constant $R(k,n,N)$ if 
\[ R(k,n,N) := \min_{c\ge0}\,c\,\,\,\mbox{subject to}\,\,\,(1-c)\|x\|_2^2 \le \|Ax\|_2^2 \le (1+c) \|x\|_2^2,\,\,\forall x\in\chi^N(k).\]
\item
Observe that $\chi^N(k)\subset\chi^N(k+1)$ for any $k$ and therefore the constants $L(k,n,N)$, $U(k,n,N)$, and $R(k,n,N)$ are nondecreasing in $k$ \cite{CTdecoding}.  
\item
For all expressions involving $L(\cdot,n,N)$ it is understood, without explicit statement, that the first argument is limited to the range where $L(\cdot,n,N)<1$.  Beyond this range of sparsity, there exist vectors which are mapped to zero, and are unrecoverable.
\end{enumerate}
\end{remark}

Using the aRIP, we analyze the three algorithms in the case of a general measurement
matrix $A$ of size $n\times N$.  For each algorithm, the application
of Definition~\ref{def:LU} results in a relaxation of the conditions
imposed on $A$ to provably guarantee recovery of all $x\in\chi^N(k)$.
We first present a stability result for each algorithm in
terms of bounding the approximation error of the output after $l$
iterations. The bounds show a multiplicative stability constant in
terms of aRIP contants that amplifies the total energy of the noise.
As a corollary, we obtain a  sufficient condition on $A$ in terms of
the aRIP for exact recovery of all $k$-sparse vectors.
The proofs of these results are found in the Appendix.  These theorems and corollaries take the same form, differing for each algorithm only by the formulae for various factors.  We state the general form of the theorems and corollaries, analogous to Theorem \ref{thm:alg} and Corollary \ref{cor:alg}, and then state the formulae for each of the algorithms CoSaMP, SP, and IHT.

\begin{thm}\label{thm:alg_knN}
Given a matrix $A$ of size $n\times N$ with aRIP constants
$L(\cdot,n,N)$ and $U(\cdot,n,N)$, for  any $x\in\chi^N(k)$,
let $y=Ax+e$, for some (unknown) noise vector $e$. Then there exists
$\mu^{alg}(k,n,N)$ such that if $\mu^{alg}(k,n,N)<1$, the output $\hat{x}$ of algorithm ``alg'' at the $l^{th}$ iteration approximates $x$ within the bound
\begin{equation}\label{eq:alg_knN}
\|x-\hat{x}\|_2 \le \kappa^{alg}(k,n,N)\left[\mu^{alg}(k,n,N)\right]^l \|x\|_2 + \frac{\xi^{alg}(k,n,N)}{1-\mu^{alg}(k,n,N)}\|e\|_2,
\end{equation}  
for some  $\kappa^{alg}(k,n,N)$ and $\xi^{alg}(k,n,N)$.
\end{thm}

\begin{cor}\label{cor:alg_knN}
Given a matrix $A$ of size $n\times N$ with aRIP constants
$L(\cdot,n,N)$ and $U(\cdot,n,N)$, for  any $x\in\chi^N(k)$,
let $y=Ax$. Then there exists
$\mu^{alg}(k,n,N)$ such that if $\mu^{alg}(k,n,N)<1$, 
 the algorithm ``alg''
exactly recovers $x$ from $y$ and $A$ in a finite number of iterations not to exceed 
\begin{equation}
\ell^{alg}_{max}(x):=\left\lceil\frac{\log\nux-\log\kappa^{alg}(k,n,N)}{\log\mu^{alg}(k,n,N)}\right\rceil +1
\end{equation}
where $\nux$ defined as in \eqref{eq:numin}.
\end{cor}

We begin with Algorithm~\ref{alg:CSP}, the Compressive Sampling
Matching Pursuit recovery algorithm of Needell and Tropp
\cite{NeTr09_cosamp}.  We relax the sufficient recovery condition in \cite{NeTr09_cosamp} via
the aRIP.  

\begin{thm}[CoSaMP]\label{thm:CSPnoisy}
Theorem \ref{thm:alg_knN} and Corollary \ref{cor:alg_knN} are satisfied by CoSaMP, Algorithm \ref{alg:CSP}, with 
$\kappa^{csp}(k,n,N):=1$ and $\mcsp$ and $\xicsp$ defined as
\begin{equation}
\mcsp:= \frac{1}{2}\left(2+\frac{L(4k,n,N)+U(4k,n,N)}{1-L(3k,n,N)}\right)\left(\frac{L(2k,n,N)+U(2k,n,N)+L(4k,n,N)+U(4k,n,N)}{1-L(2k,n,N)}\right)\label{eq:CSPCk}\\
\end{equation}
and
\begin{equation}\label{eq:CSPxi}
\xicsp:=2\left\{\left(2+\frac{L(4k,n,N)+U(4k,n,N)}{1-L(3k,n,N)}\right)\left(\frac{\sqrt{1+U(2k,n,N)}}{1-L(2k,n,N)}\right)+\frac{1}{\sqrt{1-L(3k,n,N)}}\right\}.
\end{equation}
\end{thm}

Next, we apply the aRIP to Algorithm~\ref{alg:SP}, Dai and Milenkovic's Subspace Pursuit \cite{SubspacePursuit}.  Again, the aRIP provides a sufficient condition that admits a wider range of measurement matrices than admitted by the symmetric RIP condition derived in \cite{SubspacePursuit}.

\begin{thm}[SP]\label{thm:SPnoisy}
Theorem \ref{thm:alg_knN} and Corollary \ref{cor:alg_knN} are satisfied by Subspace Pursuit, Algorithm \ref{alg:SP}, with $\kappa^{sp}(k,n,N)$, $\msp$, and $\xispk$ defined as 
\begin{align}
\kappa^{sp}(k,n,N) &:= 1+\frac{U(2k,n,N)}{1-L(k,n,N)},  \label{eq:SPkappak} \\
\msp &:= \frac{2 U(3k,n,N)}{1-L(k,n,N)} \left(1+\frac{2U(3k,n,N)}{1-L(2k,n,N)} \right)\left(1+ \frac{U(2k,n,N)}{1-L(k,n,N)}\right)\label{eq:SPmuk}
\end{align}
and
\begin{align}
\xispk & :=\frac{\sqrt{1+U(k,n,N)}}{1-L(k,n,N)}\left[1-\msp+2\kappa^{sp}(k,n,N)\left(1+\frac{2U(3k,n,N)}{1-L(2k,n,N)}\right)\right]\nonumber\\
&+\frac{2\kappa^{sp}(k,n,N)}{\sqrt{1-L(2k,n,N)}}.\label{eq:SPxi}
\end{align}
\end{thm}

Finally, we apply the aRIP analysis to Algorithm~\ref{alg:IHT},
Iterative Hard Thresholding for Compressed Sensing introduced by
Blumensath and Davies \cite{BlDa08_iht}.  Theorem~\ref{thm:IHTnoisy}
employs the aRIP to provide a weaker sufficient condition than derived
in \cite{BlDa08_iht}.

\begin{thm}[IHT]\label{thm:IHTnoisy}
Theorem \ref{thm:alg_knN} and Corollary \ref{cor:alg_knN} are satisfied by Iterative Hard Thresholding, Algorithm \ref{alg:IHT}, with $\kappa^{iht}(k,n,N):=1$ and 
$\miht$ and $\xiiht$ defined as
\begin{equation}
\miht:=2\sqrt{2}\max\left\{\omega\left[1+U(3k,n,N)\right]-1,1-\omega\left[1-L(3k,n,N)\right]\right\}.\label{eq:IHTCk}\\
\end{equation}
and
\begin{equation}\label{eq:IHTxi}
\xiiht:=2\omega\sqrt{1+U(2k,n,N)}.
\end{equation}
\end{thm}

\begin{remark}\label{rem:aRIPalg}
Each of Theorems \ref{thm:CSPnoisy}, \ref{thm:SPnoisy} and
\ref{thm:IHTnoisy} are derived following the same recipe as in
\cite{NeTr09_cosamp}, \cite{SubspacePursuit} and \cite{BlDa08_iht},
respectively, using the aRIP rather than the RIP and taking care to
maintain the least restrictive bounds at each step 
(for details, see the Appendix).  For Gaussian matrices, the aRIP
improves the lower bound on the phase transitions by nearly a multiple
of 2 when compared to similar statements using the classical RIP.  For
IHT, the aRIP is simply a scaling of the matrix so that its RIP bounds
are minimal.  This is possible for IHT as the factors in $\miht$ involve $L(\alpha k,n,N)$ and $U(\alpha k,n,N)$ for only one value of $\alpha$, here $\alpha=3$.  No such scaling interpretation is possible for CoSaMP and SP.
\end{remark}

At this point, we digress to mention that the first greedy algorithm
shown to have guaranteed exact recovery capability is Needell and
Vershynin's ROMP (Regularized Orthogonal Matching Pursuit)
\cite{NeVe06_UUP}.  We omit the algorithm and a rigorous discussion of the result, but state an aRIP condition that will guarantee sparse recovery.  ROMP chooses additions to the approximate support sets at each iteration with a regularization step requiring comparability between the added terms.  This comparability requires a proof of partitioning a vector of length $N$ into subsets with comparable coordinates,  namely the magnitudes of the elements of the subset differ by no more than a factor of 2.  The proof that such a partition exists, with each partition having a nonzero energy, forces a pessimistic bound that decays with the problem size.
\begin{thm}[Regularized Orthogonal Matching Pursuit]\label{thm:ROMP}
Let $A$ be a matrix of size $n\times N$ with aRIP constants $L(2k,n,N)$ and $U(2k,n,N)$.  Define
\begin{equation}
\mromp := U(2k,n,N)\left(1+\frac{1+U(2k,n,N)}{1-L(2k,n,N)}\right).\label{eq:RIPRomp}\\
\end{equation}
If $\mromp <\left(1+\sqrt{\frac{5n}{n-1}(\log n + 2)}\right)^{-1}$, then ROMP is guaranteed to exactly recover any $x\in\chi^N(k)$ from the measurements $y=Ax$ in a finite number of iterations.
\end{thm}

Unfortunately, this dependence of the bound on the size of the problem instance forces the result to be inadequate for large problem instances.  In fact, this result is inferior to the results for the three algorithms stated above which are all independent of problem size and therefore applicable to the most interesting cases of compressed sensing, when $(k,n,N)\rightarrow\infty$ and $\d=n/N\rightarrow 0$.  It is possible that this dependence on the problem size is an artifact of the technique of proof; without removing this dependence, large problem instances will require the measurement matrix to be a true isometry and the phase transition framework of the next section does not apply.

\section{Phase Transitions for Greedy Algorithms with Gaussian Matrices}\label{sec:PhaseTrans}

The quantities $\mu^{alg}(k,n,N)$ and $\xi^{alg}(k,n,N)$ in Theorems
\ref{thm:CSPnoisy}, \ref{thm:SPnoisy}, and \ref{thm:IHTnoisy} dictate
the current theoretical convergence bounds for CoSaMP, SP, and IHT.
Although some comparisons can be made between the forms of $\mu^{alg}$
and $\xi^{alg}$ for different algorithms, it is not possible to
quantitatively state for what range of $k$ the algorithm will satisfy
bounds on $\mu^{alg}(k,n,N)$ and $\xi^{alg}(k,n,N)$ for a specific
value of $n$ and $N$.  To establish quantitative interpretations of
the conditions in Theorems
\ref{thm:CSPnoisy}, \ref{thm:SPnoisy} and \ref{thm:IHTnoisy}, it is
necessary to have quantitative bounds on the behaviour of the aRIP
constants $L(k,n,N)$ and $U(k,n,N)$ for the matrix $A$ in question,
\cite{BCT09_DecayRIP, RIconstants}.
Currently, there is no known matrix $A$ for which it has been proven
that $U(k,n,N)$ and $L(k,n,N)$ remain bounded above and away from one,
respectively, as $n$ grows,
for $k$ and $N$ proportional to $n$.  However, it is known that for
some random matrix ensembles, with exponentially high probability on
the draw of $A$, $\frac{1}{1-L(k,n,N)}$ and $U(k,n,N)$ do remain bounded
as $n$ grows, for $k$ and $N$ proportional to $n$.  The ensemble with
the best known bounds on the growth rates of 
$L(k,n,N)$ and $U(k,n,N)$ in this setting is the Gaussian
ensemble.  
In this section, we consider large problem sizes as
$(k,n,N)\rightarrow\infty$, with $\nN\rightarrow\d$ and
$\kn\rightarrow\r$ for $\d,\r\in(0,1)$. 
We study the implications of the sufficient conditions from Section
\ref{sec:ARIP} for matrices with Gaussian i.i.d.\ entries, namely, entries drawn i.i.d.\ from the normal
distribution with mean $0$ and variance $n^{-1}$,
$\mathcal{N}(0,n^{-1})$.

Gaussian random matrices are well studied and much is known about the
behavior of their eigenvalues.  Edelman \cite{EdelmanEigenvalues88} derived bounds on the probability distribution functions of the largest and smallest eigenvalues of the Wishart matrices derived from a matrix $A$ with Gaussian i.i.d.\ entries.  Select a subset of columns indexed by $I\subset\{1,\dots,N\}$ with cardinality $k$ and form the submatrix $A_I$, and the associated Wishart matrix derived from $A_I$ is the matrix $A_I^*A_I$. The distribution of the most extreme eigenvalues of all ${N\choose k}$ Wishart matrices derived from $A$ with Gaussian i.i.d.\ entries is only of recent interest and the exact probability distribution functions are not known.  Recently, using Edelman's bounds \cite{EdelmanEigenvalues88}, the first three authors \cite{RIconstants} derived upper bounds on the probability distribution functions for the most extreme eigenvalues of all ${N\choose k}$ Wishart matrices derived from $A$.  These bounds enabled them to formulate upper bounds on the aRIP constants, $L(k,n,N)$ and $U(k,n,N)$,  for matrices of size $n\times N$ with Gaussian i.i.d.\ entries.

\begin{figure}[t]
 
 \begin{center}
 \includegraphics[bb= 70 215 546 589, width=2.70 in,height=2.00 in]{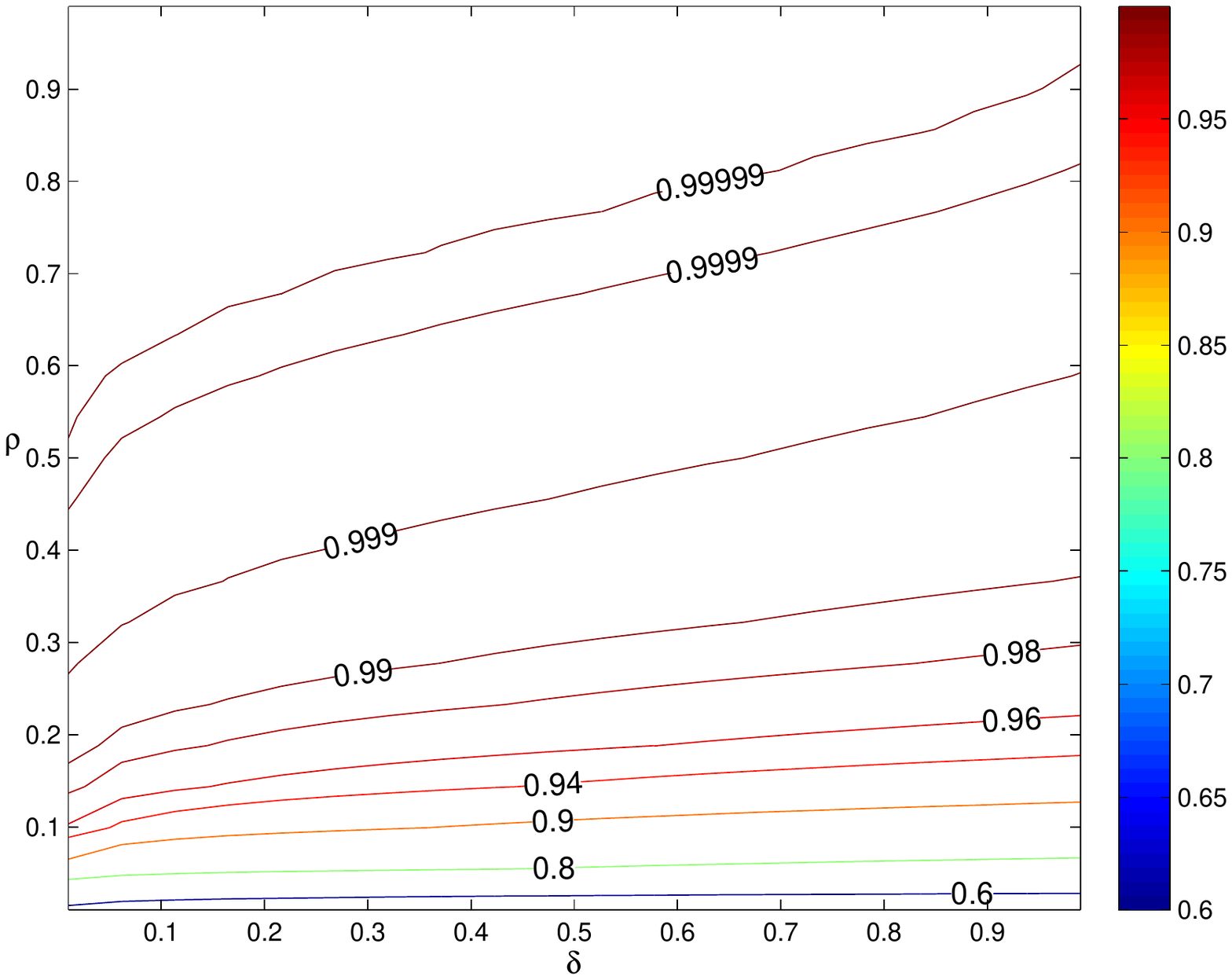}
\includegraphics[bb= 70 215 546 589, width=2.70 in,height=2.00 in]{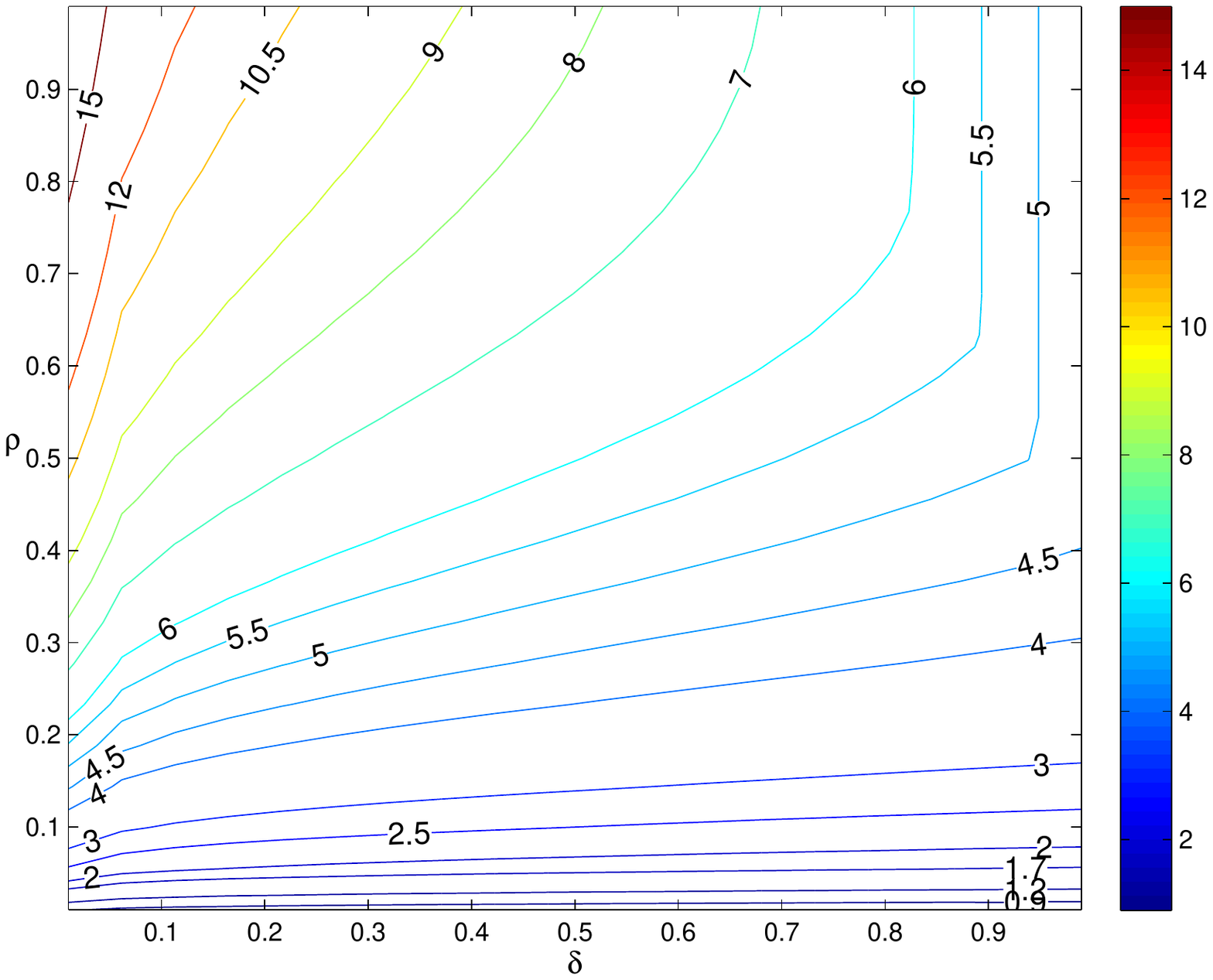}
\end{center}

 \caption{Bounds, $L(\d,\r)$ and $U(\d,\r)$ (left and right
   respectively), above which it is exponentially unlikely that the
   RIP constants $L(k,n,N)$ and $U(k,n,N)$ exceed, with entries in $A$
 drawn i.i.d. $N(0,n^{-1})$ and in the limit as $\kn\rightarrow\r$
 and $\nN\rightarrow\d$ as $n\rightarrow\infty$, see Theorem \ref{thm:LUbounds}.\label{fig:LU}}
 \end{figure}

\begin{thm}[Blanchard, Cartis, and Tanner \cite{RIconstants}]
Let $A$ be a matrix of size $n\times N$ whose entries are drawn i.i.d.\ from 
${\cal N}(0,n^{-1})$ and let $n\rightarrow\infty$ with $\kn\rightarrow\r$ and $\nN\rightarrow\d$.  Let $H(p):=p\log(1/p)+(1-p)\log(1/(1-p))$ denote the usual Shannon 
Entropy with base $e$ logarithms, and let
\begin{eqnarray}
\psi_{min}(\lambda,\rho) & := & H(\rho)+\frac{1}{2}\left[ 
(1-\rho)\log\lambda +1-\rho+\rho\log\rho-\lambda \right],\label{eq:psimin} \\
\psi_{max}(\lambda,\rho) & := & \frac{1}{2}\left[
(1+\rho)\log\lambda +1+\rho-\rho\log\rho-\lambda \right].\label{eq:psimax} 
\end{eqnarray}
Define $\lambda_{min}(\d,\r)$ and 
$\lambda_{max}(\d,\r)$ as the solution to \eqref{eq:lmin} 
and \eqref{eq:lmax}, respectively:
\begin{equation}
\delta\psi_{min}(\lambda_{min}(\d,\r),\rho)+H(\rho\delta)=0 
\quad\mbox{ for }\quad \lambda_{min}(\d,\r)\le 1-\r\label{eq:lmin}
\end{equation}
\begin{equation}
\delta\psi_{max}(\lambda_{max}(\d,\r),\rho)+H(\rho\delta)=0
\quad\mbox{ for }\quad \lambda_{max}(\d,\r)\ge 1+\r. \label{eq:lmax}
\end{equation}
Define $L(\d,\r)$ and $U(\d,\r)$ as 
\begin{equation}
L(\d,\r):= 1-\lmin(\delta,\rho) \quad \hbox{and}\quad 
U(\d,\r):= \min_{\nu\in[\r,1]}\lmax(\d,\nu)-1. \label{eq:LUdeltarho}
\end{equation}
For any $\e>0$, as $n\rightarrow\infty$,
\begin{equation*}
\hbox{Prob}\left(L(k,n,N)<L\left(\d,\r\right)+\e\right)\rightarrow 1\quad \hbox{and}\quad \hbox{Prob}\left(U(k,n,N)<U\left(\d,\r\right)+\e\right)\rightarrow 1.
\end{equation*}\label{thm:LUbounds}
\end{thm}

The details of the proof of Theorem \ref{thm:LUbounds} are found in
\cite{RIconstants}.  The bounds are derived using a simple union bound over all $N\choose k$ of the $k\times k$ Wishart 
matrices $A_I^*A_I$ that can be formed from columns of $A$.
Bounds on the tail behavior of the probability distribution function 
for the largest and smallest eigenvalues of $A_I^*A_I$ can 
be expressed in the form $p(n,\lambda)\exp(n\psi(\lambda,\r))$ with
$\psi$ defined in \eqref{eq:psimin} and \eqref{eq:psimax} and $p(n,\lambda)$ a polynomial.
Following standard practices in large deviation analysis, the tails of the 
probability distribution functionals are 
balanced against the exponentially large number of Wishart matrices 
\eqref{eq:lmin} and \eqref{eq:lmax} to define upper and lower bounds on 
the largest and smallest eigenvalues of all $N\choose k$ Wishart matrices, 
with bounds $\lambda_{max}(\d,\r)$ and $\lambda_{min}(\d,\r)$, respectively.
Overestimation of the union bound over the combinatorial number of $N\choose k$ 
Wishart matrices causes the bound $\lambda_{max}(\d,\r)$ to not be strictly increasing in $\r$ for $\d$ large;  to utilize the best available bound on the extreme of the largest eigenvalue, we note that any bound $\lambda_{max}(\d,\nu)$ for $\nu\in [\r,1]$ is also a valid bound for submatrices of size $n\times k$.
The asymptotic bounds of the aRIP constants, $L(\d,\r)$ and 
$U(\d,\r)$, follow directly.  See Figure \ref{fig:LU} for level curves of the bounds.

With Theorem~\ref{thm:LUbounds}, we are able to formulate quantitative statements about the matrices $A$ with Gaussian i.i.d.\ entries which satisfy the sufficient aRIP conditions from Section~\ref{sec:ARIP}.  A naive replacement of each $L(\cdot,n,N)$ and $U(\cdot,n,N)$ in Theorems~\ref{thm:CSPnoisy}-\ref{thm:IHTnoisy} with the asymptotic aRIP bounds in Theorem~\ref{thm:LUbounds}  is valid in these cases.  The properties necessary for this replacement are detailed in Lemma \ref{lem:superF}, stated in the Appendix.  For each algorithm (CoSaMP, SP and IHT) the recovery conditions can be stated in the same format as Theorem \ref{thm:alg} and Corollary \ref{cor:alg}, with only the expressions for $\kappa(\d,\r)$, $\mu(\d,\r)$ and $\xi(\d,\r)$ differing.  These recovery factors are stated in Theorems \ref{thm:CSPnoisyphase}-\ref{thm:IHTnoisyphase}.

\begin{thm}\label{thm:CSPnoisyphase}
Theorem \ref{thm:alg} and Corollary \ref{cor:alg} are satisfied for CoSaMP, Algorithm \ref{alg:CSP}, with $\kappa^{csp}(\d,\r):=1$ and $\mcspdr$ and $\xicspdr$ defined as
\begin{equation}\label{eq:CSPmudr}
\mcspdr:=\frac{1}{2}\left(2+\frac{L(\d,4\r)+U(\d,4\r)}{1-L(\d,3\r)}\right)\left(\frac{L(\d,2\r)+U(\d,2\r)+L(\d,4\r)+U(\d,4\r)}{1-L(\d,2\r)}\right).
\end{equation}
and 
\begin{equation}\label{eq:CSPxidr}
\xicspdr:=2\left\{\left(2+\frac{L(\d,4\r)+U(\d,4\r)}{1-L(\d,3\r))}\right)\left(\frac{\sqrt{1+U(\d,2\r)}}{1-L(\d,2\r)}\right)+\frac{1}{\sqrt{1-L(\d,3\r)}}\right\}.
\end{equation}
\end{thm}

The phase transition lower bound $\rcsp$ is defined as the solution to $\mcspdr=1$.
$\rcsp$ is displayed as the black curve in Figure \ref{fig:transitions}(a).
$\mcspdr$ and $\xicspdr/(1-\mcspdr)$ are displayed in Figure
\ref{fig:Stab_sp_csp} panels (a) and (b) respectively.

\begin{thm}\label{thm:SPnoisyphase}
Theorem \ref{thm:alg} and Corollary \ref{cor:alg} are satisfied for Subspace Pursuit, Algorithm \ref{alg:SP}, with $\kappa^{sp}(\d,\r)$, $\mspdr$, and $\xispdr$ defined as
\begin{align}
\kappa^{sp}(\d,\r)&:= 1+\frac{U(\d,2\r)}{1-L(\d,\r)}, \label{eq:SPkappadr} \\
\mspdr&:=\frac{2 U(\d,3\r)}{1-L(\d,\r)}\left(1+\frac{2U(\d,3\r)}{1-L(\d,2\r)} \right)\left(1+ \frac{U(\d,2\r)}{1-L(\d,\r)}\right), \label{eq:SPmudr}
\end{align}
and 
\begin{align}
\xispdr  & :=\frac{\sqrt{1+U(\d,\r)}}{1-L(\d,\r)}\left[1-\mspdr+2\kappa^{sp}(\d,\r)\left(1+\frac{2U(\d,3\r)}{1-L(\d,2\r)}\right)\right]\nonumber\\
&+\frac{2\kappa^{sp}(\d,\r)}{\sqrt{1-L(\d,2\r)}}.\label{eq:SPxidr}
\end{align}
\end{thm}

The phase transition lower bound $\rsp$ is defined as the solution to $\mspdr=1$.
$\rsp$ is displayed as the magenta curve in Figure \ref{fig:transitions}(a).
$\mspdr$ and $\xispdr/(1-\mspdr)$ are displayed in Figure
\ref{fig:Stab_sp_csp} panels (c) and (d) respectively.

\begin{thm}\label{thm:IHTnoisyphase}
Theorem \ref{thm:alg} and Corollary \ref{cor:alg} are satisfied for
Iterative Hard Thresholding, Algorithm \ref{alg:IHT}, with 
$\omega:=2/(2+U(\d,3\r)-L(\d,3\r))$,
$\kappa^{iht}(\d,\r):=1$, and $\mihtdr$ and $\xiihtdr$ defined as
\begin{equation}\label{eq:IHTmudr}
\mihtdr:=2\sqrt{2}\left(\frac{L(\d,3\r)+U(\d,3\r)}{2+U(\d,3\r)-L(\d,3\r)}\right)
\end{equation}
and 
\begin{equation}\label{eq:IHTxidr}
\xiihtdr:=\frac{4\sqrt{1+U(\d,2\r)}}{2+U(\d,3\r)-L(\d,3\r)}.
\end{equation}
\end{thm}

The phase transition lower bound $\riht$ is defined as the solution to $\mihtdr=1$.
$\riht$ is displayed as the red curve in Figure \ref{fig:transitions}(a).
$\mihtdr$ and $\xiihtdr/(1-\mihtdr)$ are displayed in Figure
\ref{fig:Stab_sp_csp} panels (e) and (f) respectively.

An analysis similar to that presented here for the greedy algorithms
CoSaMP, SP, and IHT was previously carried out in \cite{RIconstants}
for the $\lone$-regularization problem \eqref{eq:l1}.  The form of the
results differs from those of Theorem \ref{thm:alg} and Corollary
\ref{cor:alg} in that no algorithm was specified for how
\eqref{eq:l1} is solved.  For this reason, no results are stated for
the convergence rate or number of iterations.  
However,  \eqref{eq:l1}  can be reformulated as a 
convex quadratic or second-order cone programming problem ---
and its noiseless variant as a linear programming ---
which have polynomial complexity
when solved using interior point methods \cite{NN}. Moreover,
convergence and complexity of other alternative algorithms for solving \eqref{eq:l1} such as
gradient projection have long been studied by the optimization
community for more general problems \cite{Bertsekas, nesterovbook, NW}, and recently, more specifically
for \eqref{eq:l1} \cite{GPSR, Nesterov} and many more.
For completeness, we
include the recovery conditions for $\lone$-regularization derived in
\cite{RIconstants}; these results follow from the original
$\lone$-regularization bound derived by Foucart and Lai
\cite{FoucartLai08} for general $A$.

\begin{thm}[Blanchard, Cartis, and Tanner \cite{RIconstants}] \label{thm:Lonenoisyphase}
Given a matrix $A$ with entries drawn  i.i.d. from
${\cal N}(0,n^{-1})$,  for any $x\in\chi^N(k)$, let 
 $y=Ax+e$ for some (unknown) noise vector $e$. Define 
\begin{equation}\label{eq:FL1}
\mu^{\lone}(\d,\r):=\frac{1+\sqrt{2}}{4}\left(\frac{1+U(\d,2\r)}{1-L(\d,2\r)} -1 \right)
\end{equation}
and 
\begin{equation}\label{eq:FL1xidr}
\xi^{\lone}(\d,\r):=\frac{3(1+\sqrt{2})}{1-L(\d,2\r)}
\end{equation}
with $L(\delta,\cdot)$ and $U(\delta,\cdot)$ defined as in Theorem
\ref{thm:LUbounds}.  Let $\rfl1$ be the unique solution to $\mu^{\lone}(\d,\r)=1$.
For any $\epsilon>0$, as $(k,n,N)\rightarrow\infty$ with
$n/N\rightarrow\d\in(0,1)$ and $k/n\rightarrow\r<(1-\e)\rfl1$, 
there is an exponentially high probability on the draw
of $A$ that 
\[
\hat{x}:=\argmin_z\|z\|_{\lone}\quad\quad\mbox{ subject to }\quad
\|Az-y\|_2\le \|e\|_2
\]
approximates $x$ within the  bound
\begin{equation}\label{eq:SPnoisyphase1}
\|x-\hat{x}\|_2 \le
\frac{\xi^{\lone}(\d,(1+\e)\r)}{1-\mu^{\lone}(\d,(1+\e)\r)}\|e\|_2.
\end{equation}  
\end{thm}

\begin{figure}[t]
 
 \begin{center}
\begin{tabular}{cc}
\includegraphics[bb= 70 215 546 589, width=2.70 in,height=2.00 in]{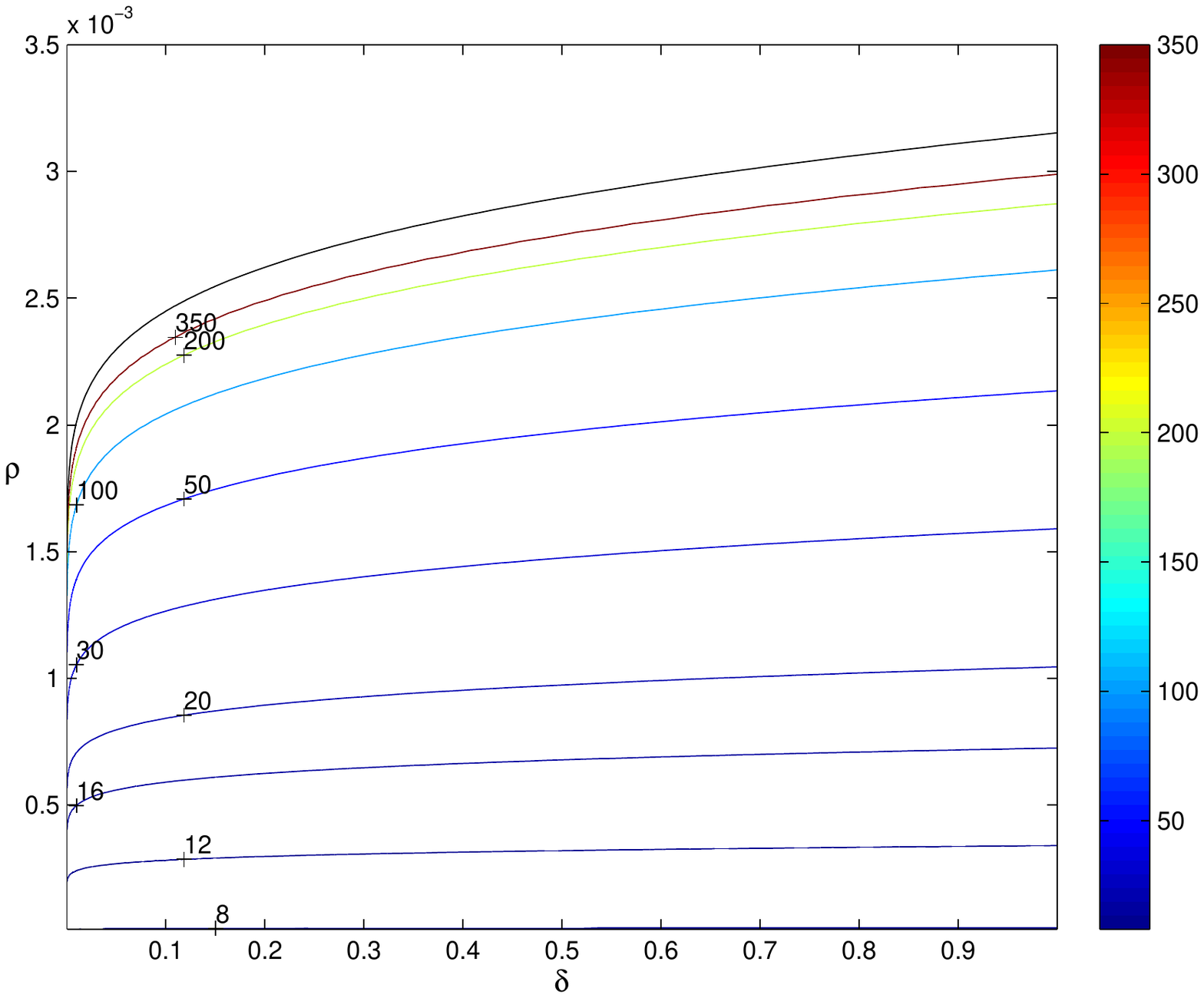} &
\includegraphics[bb= 70 215 546 589, width=2.70 in,height=2.00 in]{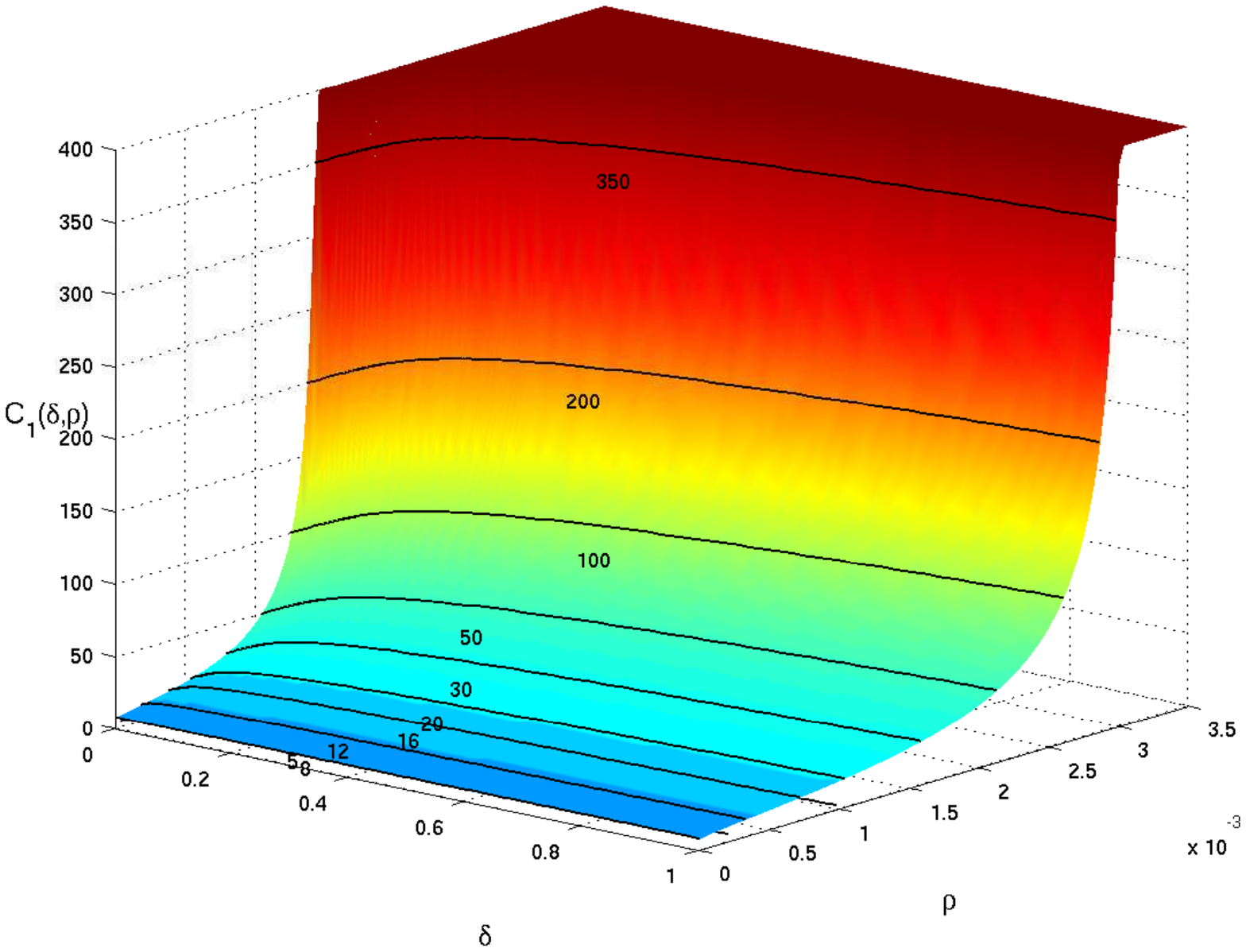} \\
(a) & (b) \\
\end{tabular}
 \caption{(a): level curves
   for specific values of $\frac{\xi}{1-\mu}(\d,\r)$ 
   for $\lone$-regularization respectively.
(b): the surface whose level curves specify the multiplicative stability constants
   for $\lone$-regularization.  \label{fig:Stab_l1_iht}}
 \end{center}
 \end{figure}

$\rfl1$ is displayed as the blue curve in Figure \ref{fig:transitions}(a).
$\mu^{\lone}(\d,\r)$ and $\xi^{\lone}(\d,\r)/(1-\mu^{\lone}(\d,\r))$
are displayed in Figure \ref{fig:Stab_l1_iht} panels (c) and (d) respectively.

\begin{cor}[Blanchard, Cartis, and Tanner \cite{RIconstants}]
\label{thm:Lonephase}
Given a matrix $A$ with entries drawn  i.i.d. from
${\cal N}(0,n^{-1})$,  for any $x\in\chi^N(k)$, let 
 $y=Ax$. For any $\epsilon>0$, with $n/N\rightarrow\d\in(0,1)$ and
$k/n\rightarrow\r<(1-\e)\rfl1$ as $(k,n,N)\rightarrow\infty$, there
is an exponentially high probability on the draw of $A$ that 
\[
\hat{x}:=\argmin_z\|z\|_{\lone}\quad\quad\mbox{ subject to }\quad
Az=y
\]
exactly recovers $x$ from $y$ and $A$.
\end{cor}

\section{Discussion and Conclusions}\label{sec:discussion}

\paragraph{Summary}
We have presented a framework in which recoverability results for
sparse approximation algorithms derived using the ubiquitous RIP can
be easily compared.  This phase transition framework,
\cite{Do05_signal,DoTa05_signal,DoTa08_JAMS,RIconstants}, translates
the generic RIP-based conditions of Theorem \ref{thm:alg_knN} into
specific sparsity levels $k$ and problem sizes $n$ and $N$ for which
the algorithm is guaranteed to satisfy the sufficient RIP conditions
with high probability on the draw of the measurement matrix; see
Theorem \ref{thm:alg}. 
Deriving (bounds on) the phase transitions
requires bounds on the behaviour of the measurement matrix' RIP
constants \cite{BCT09_DecayRIP}.
To achieve the most favorable  quantitative bounds on the phase transitions, we used the less restrictive aRIP constants; moreover, we employed the best known bounds on aRIP constants, those provided for Gaussian matrices \cite{RIconstants}, see Theorem \ref{thm:LUbounds}.

This framework was illustrated on three exemplar greedy algorithms:
CoSaMP \cite{NeTr09_cosamp}, SP
\cite{SubspacePursuit}, and IHT
\cite{BlDa08_iht}.  The lower bounds on the phase transitions in
Theorems \ref{thm:CSPnoisyphase}-\ref{thm:IHTnoisyphase} allow for a 
direct comparison of  the current theoretical results/guarantees for these algorithms.

\paragraph{Computational Cost of CoSaMP, SP
  and IHT} The major computational cost per iteration in these algorithms is the
application of one or more pseudoinverses.  
SP uses two pseudoinverses of dimensions $k\times n$ per
iteration and another to compute the output vector $\hat{x}$; see
Algorithm 2. CoSaMP
uses only one pseudoinverse per iteration but of dimensions $2k\times
n$; see Algorithm 1. Consequently, CoSaMP and SP have identical computational cost per
iteration, of order $kn^2$, if the pseudoinverse is solved using an exact
$QR$ factorization.  IHT avoids computing a pseudoinverse altogether
in internal iterations, but is aided by one pseudoinverse of
dimensions $k\times n$ on the final support set. Thus IHT has a
substantially lower computational cost per iteration than  CoSaMP and SP.
Note that pseudoinverses may be computed approximately by an iterative
method such as conjugate gradients \cite{NeTr09_cosamp}.  As such, the
exact application of a pseudoinverse could be entirely avoided, 
improving the implementation costs of these algorithms, especially of
CoSaMP and SP.  

Globally, all three algorithms converge linearly; in fact, they
converge in a finite number of iterations provided there exists a
$k$-sparse solution to $Ax=y$ and a sufficient aRIP condition is
satisfied, see Corollary \ref{cor:alg}.  For each algorithm,  the
upper bound on the required number of iterations grows unbounded as
the function $\mu^{alg}(k,n,N)\rightarrow1$.  Hence, according to the
bounds presented here, to ensure rapid convergence, it is
advantageous to have a matrix that satisfies a more
strict condition, such as $\mu^{alg}(k,n,N)<\frac12$.  Similarly, the
factor controlling stability to additive noise, namely the vector $e$
in Theorem \ref{thm:alg}, blows up as the function
$\mu^{alg}(k,n,N)\rightarrow1$.  Again, according to the bounds
presented here, in order to guarantee stability with small
amplification of the additive noise, it is necessary to restrict the
range of $\frac{\xi^{alg}}{1-\mu^{alg}}(k,n,N)$.  A phase transition
function analogous to the functions $\r_S^{alg}(\d)$  can be easily computed in these settings as well, resulting in curves lower than those presented in Figure~\ref{fig:transitions}(a). This is the standard trade-off of compressed sensing, where one must determine the appropriate balance between computational efficiency, stability, and minimizing the number of measurements.

\paragraph{Comparison of Phase Transitions and Constants of  Proportionality}
From Figure~\ref{fig:transitions}(a), we see that the best known lower
bounds on the phase transitions for the three greedy algorithms
satisfy the ordering $\rcsp<\rsp<\riht$ for Gaussian measurement
matrices. Therefore, we now know that, at least for Gaussian matrices,
according to existing thoery, IHT has the largest region where recovery 
for all signals can be guaranteed; the regions with similar guarantees for 
SP and CoSaMP are considerably smaller.  Moreover, IHT has a lower bound 
on its computational cost.

The phase transition bounds $\r_S^{alg}(\d)$ also allow a 
 precise comparison of the recoverability
results derived for these greedy algorithms with
 those proven for $\lone$-regularization using the aRIP, see Figure
 \ref{fig:transitions}.  Although
 \cite{NeTr09_cosamp,SubspacePursuit,BlDa08_iht} have provided 
guarantees of successful sparse recovery analogous to those for
$\lone$-regularization, the greedy algorithms place a more restrictive
aRIP condition on the suitable matrices to be used in the algorithm.
However, some of the algorithms for solving the $\lone$-regularization
problem, such as interior point methods, are, in general,
computationally more expensive 
that the greedy methods discussed in this paper, and hence attention
needs to be paid to the method of choice for solving the
$\lone$-regularization problem \cite{BeFr08, GPSR}.

The lower bounds on the phase transitions presented here can also be
read as lower bounds on the constant of proportionality in the
oversampling rate, namely, taking $n\ge(\r^{alg}_S(\d))^{-1}k$ measurements
rather than the oracle rate of $k$ measurements is sufficient if
algorithm ``alg'' is used to recover the $k$-sparse signal.
From Figure~\ref{fig:transitions}(b), it is clear that according to
the conditions presented here,  the convergence of  greedy
algorithms can only be guaranteed with substantially more measurements
than for $\lone$-regularization.  The lowest possible number of
measurements (when $n=N$ so $\d=1$) for the algorithms are as follows: $n\ge907k$ for
IHT, $n\ge3124k$ for SP, and $n\ge4923k$ for CoSaMP.  On the other hand, an aRIP
analysis of $\lone$-regularization yields that linear programming
requires $n\ge317k$.  In fact, using a geometric, convex polytopes approach,
Donoho has shown that for $\lone$-regularization, $n\ge5.9k$
is a sufficient number of measurements
\cite{RIconstants,Do05_signal,Do05_polytope} when the target signal, $x$, is exactly $k$-sparse, and the multiple 5.9 increases smoothly as noise is added \cite{HassibiXu08}.

\paragraph{Future Improvements and Conclusions}
The above bounds on greedy algorithms' phase
transitions could be  improved by further refining the
algorithms' theory, namely, deriving less strict aRIP conditions on
the measurement matrix that still ensure convergence of the algorithm;
as the latter is an active research topic, we expect such
developments to take place. The phase transition framework presented
here may also be applied to such advances.
Alternatively, 
increasing the lower bounds on the phase transitions could be expected to occur
from improving the upper bounds we employed on the aRIP constants of the Gaussian
measurement matrices, see Theorem \ref{thm:LUbounds}. However, 
extensive empirical calculations of lower estimates of aRIP constants
show the latter to be 
within a factor of $1.83$ of our proven upper bounds~\cite{RIconstants}.
During the revision of this manuscript, improved bounds on the aRIP constants for 
the Gaussian ensemble were derived \cite{BaTa10}, tightening the bound to be within $1.57$ 
of lower estimates.  However, for $\rho\approx 10^{-3}$ both bounds were already 
very sharp \cite{BaTa10}, and the resulting increase of the phase transitions shown 
here was under 0.5\%.

\appendix

\section{Proofs of Main Results}\label{sec:Proofs}

We present a framework by which RIP-based convergence results of the form presented in Theorem \ref{thm:alg_knN} can be translated into results of the form of Theorem \ref{thm:alg}; that is removing explicit dependencies on RIP constants in favour of their bounds.

The proofs of Theorems \ref{thm:CSPnoisy}, \ref{thm:SPnoisy}, and \ref{thm:IHTnoisy} rely heavily on a sequence of properties of the aRIP constants, which are summarize in Lemma \ref{lem:arip} and proven in Section \ref{subsec:technical}.   Theorems \ref{thm:CSPnoisyphase}, 
\ref{thm:SPnoisyphase}, and \ref{thm:IHTnoisyphase} follow from 
Theorems \ref{thm:CSPnoisy}, \ref{thm:SPnoisy}, and \ref{thm:IHTnoisy} and 
the form of $\mu^{alg}$ and $\xi^{alg}$ as functions of $L$ and $U$; this latter point is summarized in Lemma \ref{lem:superF} which is stated and proven in Section \ref{subsec:technical}.  
The resulting Theorems \ref{thm:CSPnoisyphase}, 
\ref{thm:SPnoisyphase}, and \ref{thm:IHTnoisyphase} can then be interpreted 
in the phase transition framework advocated by Donoho et al. 
\cite{Do05_signal,Do05_polytope,DoSt06_breakdown,DoTa05_signal,DoTs06_fast},
as we have explained in Section \ref{sec:discussion}.

The remainder of the Appendix is organized by algorithms, with each subsection first proving convergence bounds for generic aRIP bounds, followed by the Gaussian specific variants as functions of $(\d,\r)$.  For the results pertaining to $\ell_1$-regularization, the reader is directed to \cite{RIconstants}.

\subsection{Technical Lemmas}\label{subsec:technical}

Throughout the analysis of the algorithms, we repeatedly use implications of the aRIP on a matrix $A$ as outlined in Lemma \ref{lem:arip}.  This lemma has been proven in the symmetric case repeatedly in the literature;  we include the proof of the asymmetric variant for completeness.

Recall that for some index sets $I,J\subset\{1,\dots,N\}$, the restriction of a vector $x$ to the set $I$ is denoted $x_I$; i.e. $(x_I)_i=x_i$ for $i\in I$ and $(x_I)_i=0$ for $i\notin I$.  Furthermore, the submatrix of $A$ derived by selecting the columns of $A$ indexed by $I$ is denoted $A_I$. In either case, $x_{I-J}$ denoted the restriction of $x$ to the set of indices in $I$ that are not in $J$; likewise $A_{I-J}$ is the submatrix formed by columms of $A$ indexed by the set $I-J$. Finally, let $\RR^I$ denote the set of vectors in $\RR^N$ whose support is contained in $I$.

\begin{lemma}[Implications of aRIP]\label{lem:arip}
Let $I$ and $J$ be two disjoint index sets, namely $I,J\subset \left\{1,\dots,N\right\}$; $I\cap J=\O$.  
Suppose $A$ is a matrix of size $n\times N$ with aRIP
constants $L(|I|+|J|,n,N)$ and $U(|I|+|J|,n,N)$, and let
$u\in\RR^{I}$, $v\in\RR^{J}$, $y\in\RR^n$, $\w\in (0,1)$, and $Id$ the
identity matrix of appropriate size.
Then Definition \ref{def:LU} implies each of the following:
\begin{enumerate}[(i)]
\item $\left\|A_I^*y\right\|_2 \le \sqrt{1+U(|I|,n,N)}\|y\|_2$
\vskip 5pt
\item $(1-L(|I|,n,N))\|u\|_2\le\left\|A_I^*A_I u\right\|_2 \le (1+U(|I|,n,N))\|u\|_2$
\vskip 5pt
\item $\left\|A_I^\dagger y\right\|_2 \le (1-L(|I|,n,N))^{-\frac12}\|y\|_2$
\vskip 5pt
\item $\left|\left\langle A_Iu, A_Jv\right\rangle\right|\le \frac12\left(L(|I|+|J|,n,N)+U(|I|+|J|,n,N)\right)\|u\|_2\|v\|_2$
\vskip 5pt
\item $\left\|A_I^*A_J v\right\|_2 \le U(|I|+|J|,n,N)\|v\|_2$.
\vskip 5pt
\item $\left\|(Id-\omega A_I^*A_I)u\right\|_2 \le
  \max\left\{\omega(1+U(|I|,n,N))-1,1-\omega(1-L(|I|,n,N))\right\}\|u\|_2.$
\end{enumerate}
\end{lemma}

\begin{pf}
From Remark~\ref{rem:ripnondecreasing}, it is clear that the aRIP constants are nondecreasing in the first argument pertaining the sparsity level.  Therefore, $A$ must also have the aRIP constants $L(|I|,n,N)\le L(|I|+|J|,n,N)$ and $U(|I|,n,N)\le U(|I|+|J|,n,N)$.  Also, Definition \ref{def:LU} implies that the singular values of the submatrix $A_I$ are contained in the interval $[\sqrt{1-L(|I|,n,N)},\sqrt{1+U(|I|,n,N)}]$.  Thus, (i)-(iii) follow from the standard relationships between the singular values of $A_I$ and the associated matrix in (i)-(iii).

To prove (iv), let $m=|I|+|J|$.  We may assume $\left\|u\right\|_2=\|v\|_2=1$; otherwise we normalize the vectors.  Let $\a=A_Iu$ and $\b=A_Jv$.  Then, since $I\cap J=\O$, 
 \begin{align}
\left\|\a\pm \b\right\|_2^2 = \left\|A_Iu\pm A_Jv\right\| &= \left\|\left[A_I, A_J\right]\left[\begin{array}{c} u \\ \pm v \end{array}\right]\right\|_2 \label{eq:lemproofa}\\
\left\|\left[\begin{array}{c} u \\ \pm v \end{array}\right]\right\|_2^2 = \left\|u\right\|_2^2 + \left\|v\right\|_2^2 &= 2.\label{eq:lemproofb}
\end{align} 

$\left[A_I, A_J\right]$ is a submatrix of $A$ of size $n\times m$, so applying Definition \ref{def:LU} to the right most portion of \eqref{eq:lemproofa}  and invoking \eqref{eq:lemproofb}, we have
\begin{equation}\label{eq:lemproofc}
2\left(1-L(m,n,N)\right)\le\|\a\pm \b\|_2^2\le2 \left(1+U(m,n,N)\right).
\end{equation}

By polarization and \eqref{eq:lemproofc}, 
\begin{align*}
\left\langle \a,\b\right\rangle=\frac{\left\|\a+\b\right\|_2^2-\left\|\a-\b\right\|_2^2}{4}  &\le \frac{L(m,n,N)+U(m,n,N)}{2} \\ 
\hbox{and}\qquad -\left\langle \a,\b\right\rangle=\frac{\left\|\a-\b\right\|_2^2-\left\|\a+\b\right\|_2^2}{4} &\le \frac{L(m,n,N)+U(m,n,N)}{2}.
\end{align*}

Thus $|\left\langle A_Iu,A_Jv\right\rangle|=|\left\langle \a,\b\right\rangle|\le \left(L(m,n,N)+U(m,n,N)\right)/2$, establishing (iv).

Since $I\cap J=\O$, the matrix $-A_I^*A_J$ is a submatrix of $Id_{I\cup J}-A_I^*A_J$.  To establish (v), we observe that the aRIP implies that the eigenvalues of every size $n\times m$ submatrix of $A$ lie in the interval $[1-L(m,n,N), 1+U(m,n,N)]$.  Thus the eigenvalues of $Id_{I\cup J}-A_I^*A_J$ must lie in $[0,U(m,n,N)]$.  Therefore $\left\|A_I^*A_Jv\right\|_2^2=\left\|-A_I^*A_Jv\right\|_2^2$ completes the proof of (v).  

To prove (vi), note that $\|(Id-\omega A_I^* A_I)\|_2$ is bounded above
by the maximum magnitude of the eigenvalues of $(\omega A_I^* A_I-Id)$, which lie
in the interval with endpoints $\omega(1-L(|I|,n,N))-1$ and $\omega(1+U(|I|,n,N))-1$.  
\end{pf}

Theorems \ref{thm:CSPnoisyphase}, 
\ref{thm:SPnoisyphase}, and \ref{thm:IHTnoisyphase} follow from 
Theorems \ref{thm:CSPnoisy}, \ref{thm:SPnoisy}, and \ref{thm:IHTnoisy} and 
the form of $\mu^{alg}$ and $\xi^{alg}$ as functions of $L$ and $U$.
We formalize the relevant functional dependencies in the next three lemmas.

\begin{lemma}\label{lem:superF}
For some $\tau<1$, define the set
$\mathcal{Z}:=(0,\tau)^p\times(0,\infty)^q$ and let
$F:\mathcal{Z}\rightarrow \RR$ be continuously  differentiable on
$\mathcal{Z}$. Let $A$ be a Gaussian matrix of size $n\times N$ with
aRIP constants $L(\cdot,n,N),U(\cdot,n,N)$ and let
$L(\d,\cdot),U(\d,\cdot)$ be defined as in Theorem~\ref{thm:LUbounds}.
Define $1$ to be the vector of all ones, and
\begin{align}
z(k,n,N)&:=[L(k,n,N),\dots,L(pk,n,N),U(k,n,N),\dots,U(qk,n,N)] \\
z(\d,\r)&:=[L(\d,\r),\dots,L(\d,p\r),U(\d,\r),\dots,U(\d,q\r)].
\end{align}

\begin{enumerate}[(i)]
\item   Suppose, for all $t\in \mathcal{Z}$, $\left(\nabla F[t]\right)_i \ge 0$ for all $i=1,\dots,p+q$ and for any $v\in\mathcal{Z}$ we have $\nabla F[t]\cdot v > 0$.  Then for any $c\e>0$, as $(k,n,N)\rightarrow\infty$ with $\nN\rightarrow\d,\kn\rightarrow\r$, there is an exponentially high probability on the draw of the matrix $A$ that 
\begin{equation}\label{eq:intermF}
\hbox{Prob}\left(F[z(k,n,N)]<F[z(\d,\r)+1c\e]\right)\rightarrow 1 \qquad \hbox{as}\ n\rightarrow\infty.
\end{equation}
\item   Suppose, for all $t\in \mathcal{Z}$, $\left(\nabla
    F[t]\right)_i \ge 0$ for all $i=1,\dots,p+q$ and there exists
  $j\in\{1,\dots,p\}$ such that $\left(\nabla F[t]\right)_j > 0$.
  Then there exists $c\in (0,1)$ depending only on $F,\d$,and $\r$
  such that for any $\e\in (0,1)$ 
\begin{equation}\label{eq:Fdr}
F[z(\d,\r)+1c\e]<F[z(\d,(1+\e)\r)],
\end{equation}
and so there is an exponentially high probability on the draw of $A$ that
\begin{equation}\label{superF:final}
\hbox{Prob}\left(F[z(k,n,N)]<F[z(\d,(1+\e)\r)]\right)\rightarrow 1 \qquad \hbox{as}\ n\rightarrow\infty.
\end{equation}
Also, $F(z(\d,\r))$ is strictly increasing in $\rho$. 
\end{enumerate}
\end{lemma}

\begin{pf}
To prove (i), suppose $u,v\in\mathcal{Z}$ with $v_i>u_i$ for all
$i=1,\dots,p+q$.  From Taylor's Theorem, $F[v]=F[u+(v-u)]=F[u]+\nabla
F[t]\cdot [v-u]$ with $t=u+\lambda[v-u]$ for some
$\lambda\in(0,1)$. Then 
\begin{equation}\label{eq:increasingC}
F[v]>F[u]
\end{equation}
since, by assumption,  $\nabla F[t]\cdot [v-u] > 0$.  

From Theorem~\ref{thm:LUbounds}, for any $c\e>0$ and any $i=1,\dots,p+q$, as $(k,n,N)\rightarrow\infty$ with $\nN\rightarrow\d$, $\kn\rightarrow\r$,
\[ \hbox{Prob}\left(z(k,n,N)_i<z(\d,\r)_i+c\e\right)\rightarrow 1, \]
with convergence to $1$ exponential in $n$.
Therefore, letting $v_i:=z(\d,\r)_i+c\e$ and $u_i:=z(k,n,N)_i$, for
all $i=1,\dots,p+q$,
we conclude from \eqref{eq:increasingC} that 
\[
\hbox{Prob}(F[z(k,n,N)]<F[z(\d,\r)+1c\e])\rightarrow 1,
\]
again with convergence to $1$ exponential in $n$.

To establish (ii), we take the Taylor expansion of $F$ centered at $z(\d,\r)$, namely
\begin{align}
F[z(\d,\r)+1c\e]&=F[z(\d,\r)]+\nabla F[t_1]\cdot 1c\e \quad \hbox{for}\ t_1\in (z(\d,\r),z(\d,\r)+1c\e)\\
F[z(\d,(1+\e)\r)]&= F[z(\d,\r)]+\left.\left(\nabla F[z(\d,\r)]\cdot \frac{\partial}{\partial\r}z(\d,\r)\right)\right|_{\r=t_2}\e\r \quad \hbox{for}\ t_2\in (\r,(1+\e)\r).
\end{align}
Select
\begin{align*}
t_1^\star&=\hbox{argmax}\left\{\nabla F[t_1] : t_1\in [z(\d,\r),z(\d,\r)+1]\right\} \\
t_2^\star&=\hbox{argmin}\left\{\left.\left(\nabla F[z(\d,\r)]\cdot \frac{\partial}{\partial\r}z(\d,\r)\right)\right|_{\r=t_2} : t_2\in [\r,(1+\e)\r]\right\}
\end{align*}
 so that 
\begin{align}
F[z(\d,\r)+1c\e]&\le F[z(\d,\r)]+\nabla F[t_1^\star]\cdot 1c\e  \label{eq:Fdr1} \\
F[z(\d,(1+\e)\r)] &\ge F[z(\d,\r)]+\left.\left(\nabla F[z(\d,\r)]\cdot \frac{\partial}{\partial\r}z(\d,\r)\right)\right|_{\r=t_2^\star}\e\r. \label{eq:Fdr2}
\end{align}
Since $L(\d,\r)$ is strictly increasing in $\r$ \cite{RIconstants},
then
$\left(\left.\frac{\partial}{\partial\r}z(\d,\r)\right|_{\r=t_2^\star}\right)_j>0$
for all $j=1,\dots,p$.  Since $U(\d,\r)$ is nondecreasing in $\r$
\cite{RIconstants}, then
$\left(\left.\frac{\partial}{\partial\r}z(\d,\r)\right|_{\r=t_2^\star}\right)_i\ge0$ for all $i=p+1,\dots,p+q$.
Hence, by the hypotheses of (ii), 
\begin{align*}
\left.\left( \nabla F[z(\d,\r)]\cdot \frac{\partial}{\partial\r}z(\d,\r)\right)\right|_{\r=t_2^\star}&>0  \\
 \nabla F[t_1^\star]\cdot 1 &> 0.
\end{align*}
Therefore, for any $c$ satisfying
\[ 0 < c < \min\left\{1,\r\frac{\left.\left(\nabla F[z(\d,\r)]\cdot\frac{\partial}{\partial\r}z(\d,\r)\right)\right|_{\r=t_2^\star}}{\nabla F[t_1^\star]\cdot 1}\right\},\]
\eqref{eq:Fdr1} and \eqref{eq:Fdr2} imply \eqref{eq:Fdr}. Since the
hypotheses of (ii) imply those of (i),  \eqref{eq:intermF}
also holds, and so \eqref{superF:final} follows. $F(z(\d,\r))$
strictly
increasing follows from the hypotheses of (ii) and $L(\d,\r)$ and
$U(\d,\r)$ strictly increasing and nondecreasing in $\r$, respectively \cite{RIconstants}.
\end{pf}

Let the superscript $alg$ denote the algorithm identifier so that
$\mu^{alg}(k,n,N)$ is defined by one of \eqref{eq:CSPCk},
\eqref{eq:SPmuk}, \eqref{eq:IHTCk}, while $\mu^{alg}(\d,\r)$ is
defined by one of \eqref{eq:CSPmudr}, \eqref{eq:SPmudr},
\eqref{eq:IHTmudr}.   Next, a simple property is summarized in Lemma
\ref{lem:levelcurveALT}, that further reveals some necessary ingredients of
our analysis.

\begin{lemma}\label{lem:levelcurveALT}  Assume that $\mu^{alg}(\d,\r)$ is
  strictly increasing in $\r$ and let $\r_S^{alg}(\d)$ solve
  $\mu^{alg}(\d,\r)=1$. For any $\e\in(0,1)$, if
  $\r<(1-\e)\r_S^{alg}(\d)$, then $\mu^{alg}(\d,(1+\e)\r)<1$. 
\end{lemma}

\begin{pf}
Let $\r^{alg}_\e(\d)$ be the solution to
$\mu^{alg}(\d,(1+\e)\r)=1$. Since by definition, 
 $\r_S^{alg}(\d)$ denotes a solution to $\mu^{alg}(\d,\r)=1$, and
 this solution is unique as $\mu^{alg}(\d,\r)$ is strictly increasing,
we must have $(1+\e)\r^{alg}_\e(\d)=\r_S^{alg}(\d)$.  Since $(1-\e)<(1+\e)^{-1}$ for all $\e\in(0,1)$, we have $(1-\e)\r_S^{alg}(\d)<\r_\e^{alg}(\d)$.  If $\r<(1-\e)\r_S^{alg}(\d)$, then since $\mu^{alg}(\d,\r)$ is strictly increasing in $\r$,
\[ \mu^{alg}(\d,(1+\e)\r)<\mu^{alg}(\d,(1+\e)(1-\e)\r_S^{alg}(\d))<\mu^{alg}(\d,(1+\e)\r_\e^{alg}(\d))=1.\]
\end{pf}

Note that Lemma \ref{lem:superF} ii) with $F:=\mu^{alg}$ will be
employed to show the first assumption in Lemma
\ref{lem:levelcurveALT}; this is but one of several good uses of 
Lemma \ref{lem:superF} that we will make.

Corollaries \ref{cor:alg} and \ref{cor:alg_knN} are easily derived
from Lemma~\ref{lem:iterations}.  
Note that this lemma demonstrates only that the support set has been recovered.   The proof of Lemma~\ref{lem:iterations} is a minor generalization of a proof from \cite[Theorem 7]{SubspacePursuit}.   

\begin{lemma}\label{lem:iterations}
Suppose, after $l$ iterations, algorithm $alg$ returns the $k$-sparse approximation $\hat{x}^l$ to a $k$-sparse target signal $x$.  Suppose there exist constants $\mu$ and $\kappa$ independent of $l$ and $x$ such that 
\begin{equation}\label{eq:iterations}
\|x-\hat{x}^l\|_2 \le \kappa \mu^l \|x\|_2.
\end{equation}
If $\mu<1$, then the support set of $\hat{x}^l$ coincides with the support set of $x$ after at most $\ell_{max}^{alg}(x)$ iterations, where
\begin{equation}\label{eq:iterations2}
\ell_{max}^{alg}(x):=\left\lceil \frac{\log
    \frac{1}{\kappa}\nu_{min}(x)}{\log \mu} \right\rceil + 1,
\end{equation}
where $\nux$ is defined in \eqref{eq:numin}.
\end{lemma}

\begin{pf}  Let $T$ be the support set of $x$ and $T^l$ be the support set of $\hat{x}^l$; as $x,\hat{x}^l\in\chi^N(k)$, $|T|,|T^l|\le k$.
From the definition \eqref{eq:iterations2} of $\ell_{max}^{alg}(x)$
and \eqref{eq:numin}, $\kappa \mu^{\ell_{max}^{alg}(x)}
\|x\|_2<\min_{i\in T}|x_i|$.  
From \eqref{eq:iterations}, we then have
\[
\|x-\hat{x}^{\ell_{max}^{alg}(x)}\|_2 \le \kappa \mu^{\ell_{max}^{alg}(x)} \|x\|_2 <\min_{i\in T}|x_i|
\]
which clearly implies that $T\subset T^{\ell_{max}^{alg}(x)}$.  Since $|T|=|T^{\ell_{max}^{alg}(x)}|$, the sets must be equal.
\end{pf}

Theorems~\ref{thm:CSPnoisy}, \ref{thm:SPnoisy} and \ref{thm:IHTnoisy}
define the constants $\mu=\mu^{alg}(k,n,N)$ and $\kappa$ to be used in
Lemma~\ref{lem:iterations} for proving Corollary \ref{cor:alg_knN}. 
For CoSaMP and IHT, $\kappa=\kappa^{alg}(k,n,N)=1$.  For SP, the term involving $\kappa$ is removed by  combining Lemmas~\ref{prop:SPnoisyA} and \ref{prop:SPnoisyB} (with $e=0$)  to obtain
\begin{equation}\label{eq:SPmu}
\|x_{T-T^l}\|_2\le \msp \|x_{T-T^{l-1}}\|_2;
\end{equation}
applying \eqref{eq:SPmu} iteratively provides
\begin{equation}\label{eq:SPmul}
\|x_{T-T^l}\|_2\le \msp^l \|x\|_2.
\end{equation}
which again, gives $\kappa=1$.  
Similarly, Theorems~\ref{thm:CSPnoisyphase}, \ref{thm:SPnoisyphase}
and \ref{thm:IHTnoisyphase} define the constants
$\mu=\mu^{alg}(\d,\r)$ and $\kappa$ to be used in
Lemma~\ref{lem:iterations} for proving Corollary \ref{cor:alg}, with the above comments on the IHT choice
of $\kappa$ also applying in this case. 

To ensure exact recovery of the target signal, namely, to complete the
proof of Corollaries \ref{cor:alg} and \ref{cor:alg_knN}, we actually need something stronger than recovering the support set as implied by Lemma~\ref{lem:iterations}.  For CoSaMP and SP, since the algorithms employ a pseudoinverse at an appropriate step, the output is then the exact sparse signal.  For IHT, no pseudoinverse has been applied; thus, to recover the signal exactly, one simply determines $T$ from the output vector and then $x=A_T^\dagger y$. 
These comments and Lemma \ref{lem:iterations} now establish
Corollaries \ref{cor:alg} and \ref{cor:alg_knN} for each algorithm, and
we will not restate the proof for each individual algorithm.

In each of the following subsections, we first consider the case of general measurement matrices, $A$, and prove the results from Section \ref{sec:ARIP} which establish an aRIP condition for an algorithm.  We then proceed to choose a specific matrix ensemble, matrices with Gaussian i.i.d.\ entries, for which Section \ref{sec:PhaseTrans} establishes lower bounds on the phase transition for exact recovery of all $x\in\chi^N(k)$ and then provide probabilistic bounds on the multiplicative stability factors.

\def\rcspeps{\rho_\epsilon^{csp}(\delta)}
\def\mcspeps{\mu^{csp}(\d,[1+\epsilon]\r)}
\newcommand{\aaa}{\ast}
\newcommand{\RRR}{\mathrm{I\!R\!}}
\newcommand{\xte}{\tilde x}
\newcommand{\xtl}{\tilde x_{T^l}}
\newcommand{\xtm}{\tilde x_{T^{l-1}}}
\newcommand{\ytl}{\tilde y^l}
\newcommand{\tlt}{\tilde T}

\subsection{Proofs for CoSaMP}\label{sec:ProofsCSMP}

In this section we prove the results from Sections \ref{sec:ARIP} and
\ref{sec:PhaseTrans} reported for CoSaMP \cite{NeTr09_cosamp}.  The
proofs mimic those of Needell and Tropp while employing the 
aRIP constants.  In each proof, the smallest possible support is
retained for the aRIP constants in order to acquire from this method
of analysis the best possible conditions on the measurement matrix
used in the CoSaMP algorithm. This change is in many
cases straightforward, requiring only a substitution of
$U(ak,n,N)$ or $L(ak,n,N)$ for $R(ak,n,N)$, for some
$a\in\{2,3,4\}$. In such cases we simply restate the result. Where
there is a more substantial change, we
provide fuller details of the proof. 

The argument proceeds in~\cite{NeTr09_cosamp} by establishing bounds on the
approximation error at a given iteration in terms of the
approximation error at the previous
iteration, and the energy of the noise. Since each iteration of the CoSaMP algorithm consists of essentially four
steps, this was achieved by a series of four
lemmas \cite[Lemmas 4.2 to 4.5]{NeTr09_cosamp}, one for each step. We
restate \cite[Lemmas 4.3 and 4.5]{NeTr09_cosamp} (the
support merger and pruning steps respectively) without any alteration,
and provide an outline of how the proofs of \cite[Lemmas 4.2 and 4.4]{NeTr09_cosamp}
(identification and estimation) can be adapted. To simplify the
working, we follow~\cite{NeTr09_cosamp} and introduce some further
notation: let the set of $2k$ indices corresponding to the largest magnitude entries
of $A^{\ast}y^{l-1}$ in Step 1 of Algorithm 1 be denoted by
$\Omega$. Also let $r=\xtm-x$ be the error in the approximation from the
previous iteration, and let $R$ be the support set of $r$, so that
$|R|\le 2k$.\\
\begin{lemma}\label{lem:CSPid}
After the identification step, we have
$$\|r_{\Omega^C}\|_2\le\left(\frac{L(2k,n,N)+U(2k,n,N)+L(4k,n,N)+U(4k,n,N)}{2(1-L(2k,n,N))}\right)\|r\|_2+2\frac{\sqrt{1+U(2k,n,N)}}{1-L(2k,n,N)}\|e\|_2.$$
\end{lemma}
\begin{pf}
By Lemma~\ref{lem:arip}, we have
$$\| y^l_{(\Omega-R)}\|_2\le\frac{1}{2}(L(4k,n,N)+U(4k,n,N))\|r\|_2+\sqrt{1+U(2k,n,N)}\|e\|_2,\;\;\;\mbox{and}$$
$$\| y^l_{(R-\Omega)}\|_2\geq(1-L(2k,n,N))\|r_{(R-\Omega)}\|_2-\frac{1}{2}(L(2k,n,N)+U(2k,n,N))\|r\|_2-\sqrt{1+U(2k,n,N)}\|e\|_2.$$
The result now follows by rearrangement.
\end{pf}

\begin{lemma}\label{lem:CSPsupp}
After the support merger step, we have
$$\|x_{(\tlt^l)^C}\|_2\le\|r_{\Omega^C}\|_2.$$
\end{lemma}

\begin{lemma}\label{lem:CSPest}
After the estimation step, we have
$$\|x-\xte\|_2\le\left(1+\frac{L(4k,n,N)+U(4k,n,N)}{2(1-L(3k,n,N))}\right)\|x_{(\tlt^l)^C}\|_2+\frac{1}{\sqrt{1-L(3k,n,N)}}\|e\|_2.$$
\end{lemma}
\begin{pf}
Using Lemma~\ref{lem:arip}, we have
$$\|x_{\tlt^l}-\xte_{\tlt^l}\|_2\le\frac{L(4k,n,N)+U(4k,n,N)}{2(1-L(3k,n,N))}\|x_{(\tlt^l)^C}\|_2+\frac{1}{\sqrt{1-L(3k,n,N)}}\|e\|_2,$$
which combines with $\|x-\xte\|_2\le\|x_{(\tlt^l)^C}\|_2+\|x_{\tlt^l}-\xte_{\tlt^l}\|_2$ to
give the required result.
\end{pf}

\begin{lemma}\label{lem:CSPprune}
After the pruning step, we have
$$\|x-\xtl\|_2\le 2\|x-\xte\|_2.$$
\end{lemma}

The preceding lemmas facilitate the proof of Theorem~\ref{thm:CSPnoisy}.

\begin{pf}
[Theorem~\ref{thm:CSPnoisy}] By Lemmas~\ref{lem:CSPest} and~\ref{lem:CSPprune}, we have
$$\begin{array}{rcl}
\|x-\xtl\|_2&\le&2\|x-\xte\|_2\\
&\le&\left(2+\frac{L(4k,n,N)+U(4k,n,N)}{1-L(3k,n,N)}\right)\|x_{(\tlt^l)^C}\|_2+\frac{2}{\sqrt{1-L(3k,n,N)}}\|e\|_2.
\end{array}$$
Now by Lemma~\ref{lem:CSPsupp}, $\|x_{(\tlt^l)^C}\|_2$ is bounded above by
$\|r_{\Omega^C}\|_2$. Then applying Lemma~\ref{lem:CSPid} and
simplifying, we obtain
\begin{equation}\label{CSPpreresult}
\|\xtl-x\|_2\le\mcsp\|\xtm-x\|_2+\xicsp\|e\|_2.
\end{equation}
Given our assumption that $\mcsp<1$, we can now prove a stronger statement, namely that for $l\geq 0$
we have
\begin{equation}\label{cspinduction}
\|\xtl-x\|_2\le\left[\mcsp\right]^l\|x\|_2+\xicsp\left(\frac{1-\left[\mcsp\right]^l}{1-\mcsp}\right)\|e\|_2.
\end{equation}
We proceed by induction. Assume the result holds for some $l\geq
0$. Then, applying the inductive hypothesis and (\ref{CSPpreresult}), we have\\
$$\begin{array}{rcl}
\|\xte_{T^{l+1}}-x\|&\le&\mcsp\|\xtl-x\|_2+\xicsp\|e\|_2\\
&\le&\mcsp\left(\left[\mcsp\right]^l\|x\|_2+\xicsp \frac{1-\left[\mcsp\right]^l}{1-\mcsp} \|e\|_2\right)+\xicsp\|e\|_2\\
&=&\left[\mcsp\right]^{l+1}\|x\|_2+\xicsp\left(\mcsp\frac{1-\left[\mcsp\right]^l}{1-\mcsp}+1\right)\|e\|_2\\
&=&\left[\mcsp\right]^{l+1}\|x\|_2+\xicsp\left(\frac{1-\left[\mcsp\right]^{l+1}}{1-\mcsp}\right)\|e\|_2,
\end{array}$$
and so the result is also true for $l+1$, and so (\ref{cspinduction}) holds for all $l\geq
0$ by induction.\\
Finally, note that 
$$\xicsp\left(\frac{1-\left[\mcsp\right]^l}{1-\mcsp}\right)\le\frac{\xicsp}{1-\mcsp}$$
for all $l\geq 0$, and also that if CoSaMP terminates after $l$ iterations we have
$\hat{x}=\xtl$.
\end{pf}

Having established the results of Section~\ref{sec:ARIP} for CoSaMP,
we now focus on Gaussian random matrices and prove the results from
Section~\ref{sec:PhaseTrans} concerning CoSAMP.

\begin{pf}[Theorem~\ref{thm:CSPnoisyphase}]
Let $x, y, A$ and $e$ satisfy the hypothesis of Theorem~\ref{thm:CSPnoisyphase} and select $\e>0$.
Fix $\tau<1$ and let
\begin{align*}
z(k,n,N)&=[L(2k,n,N), L(3k,n,N), L(4k,n,N), U(2k,n,N), U(4k,n,N)] \\
\hbox{ and }\qquad z(\d,\r)&=[L(\d,2\r), L(\d,3\r), L(\d,4\r), U(\d,2\r), U(\d,4\r)].
\end{align*}
Define $\mathcal{Z}=(0,\tau)^3\times(0,\infty)^2$ and define the functions $F^{csp}, G^{csp}:\mathcal{Z}\rightarrow\RR$:
\begin{align}
F^{csp}[z]:=F^{csp}[z_1,\dots,z_5]&= 2\left(2+\frac{z_3+z_5}{1-z_2}\right)\left(\frac{z_1+z_4+z_3+z_5}{1-z_1}\right).\label{eq:Fcsp}\\
G^{csp}[z]:=G^{csp}[z_1,\dots,z_5]&= 2\left\{\left(2+\frac{z_3+z_5}{1-z_2}\right)\left(\frac{\sqrt{1+z_4}}{1-z_1}\right)+\frac{1}{\sqrt{1-z_2}}\right\}.\label{eq:Gcsp}
\end{align}
Clearly, $\left(\nabla F^{csp}[t]\right)_i\ge0$ for all $i=1,\dots,5$ and 
\[ \left(\nabla F^{csp}[t]\right)_1=\frac12\left(2+\frac{t_3+t_5}{1-t_2}\right)\left(\frac{1+t_4+t_3+t_5}{(1-t_1)^2}\right)>0.\]
Hence the hypotheses of Lemma~\ref{lem:superF}  (ii) are satisfied for $F^{csp}$.  
By \eqref{eq:CSPCk}, \eqref{eq:CSPmudr} and \eqref{eq:Fcsp}, 
$F^{csp}[z(k,n,N)]=\mcsp$ and $F^{csp}[z(\d,\r)]=\mcspdr$.  
Thus, by Lemma~\ref{lem:superF}, as $(k,n,N)\rightarrow\infty$ with $\nN\rightarrow\d$, $\kn\rightarrow\r$, 
\begin{equation}\label{eq:FcspProb}
\hbox{Prob}\left(\mcsp<\mu^{csp}(\d,(1+\e)\r)\right)\rightarrow 1.
\end{equation}
Also, $\mcspdr$ is strictly increasing in $\r$ and so Lemma \ref{lem:levelcurveALT} applies.

Similarly, $G^{csp}$ satisfies the hypotheses of Lemma~\ref{lem:superF} (ii).
Likewise, by \eqref{eq:CSPxi}, \eqref{eq:CSPxidr} and
\eqref{eq:Gcsp}, $G^{csp}[z(k,n,N)]=\xicsp$  and $G^{csp}[z(\d,\r)]=\xicspdr$. 
Again, by Lemma~\ref{lem:superF}, as $(k,n,N)\rightarrow\infty$ with $\nN\rightarrow\d$, $\kn\rightarrow\r$, 
\begin{equation}\label{eq:GcspProb}
\hbox{Prob}\left(\xicsp<\xi^{csp}(\d,(1+\e)\r)\right)\rightarrow 1.
\end{equation}

Therefore, for any $x\in\chi^N(k)$ and any noise vector $e$, as $(k,n,N)\rightarrow\infty$ with $\nN\rightarrow\d$, $\kn\rightarrow\r$, there is an exponentially high probability on the draw of a matrix $A$ with Gaussian i.i.d.\ entries that
\begin{equation}\label{eq:cspbound}
\left[\mcsp\right]^l\|x\|_2 + \frac{\xicsp}{1-\mcsp}\|e\|_2 \le \left[\mu^{csp}(\d,(1+\e)\r)\right]^l\|x\|_2 + \frac{\xi^{csp}(\d,(1+\e)\r)}{1-\mu^{csp}(\d,(1+\e)\r)}\|e\|_2.
\end{equation} 
Combining \eqref{eq:cspbound} with Theorem~\ref{thm:CSPnoisy} completes the argument.
\end{pf}

\subsection{Proofs for Subspace Pursuit}\label{sec:ProofsSP}

In this section we outline the proofs for the results in Section
\ref{sec:ARIP} and then prove the results in Section \ref{sec:PhaseTrans} reported for SP \cite{SubspacePursuit}.  The proofs mimic those of Dai and Milenkovic while employing the aRIP constants.  In each proof, the smallest possible support is retained for the aRIP constants in order to acquire from this method of analysis the best possible conditions on the measurement matrix used in the SP algorithm.   

The index set $T$ defines the support of the target signal $x$;
$T=\hbox{supp}(x)$. For this section, the index sets $T^l,
\tilde{T}^l, T^{l\pm1}$ and the vectors $\tilde{x}, y_r^l, \hat{x}$
are defined by SP, Algorithm~\ref{alg:SP}.

We begin in the setting of an arbitrary measurement matrix $A$ of size $n\times N$ and formulate the aRIP conditions of Theorem \ref{thm:SPnoisy}.  A sequence of lemmas leads us to Theorem~\ref{thm:SPnoisy}.  Lemmas~\ref{prop:SPnoisyA} and \ref{prop:SPnoisyB} directly follow the proofs from \cite[Theorem 10]{SubspacePursuit} with the adaptation that we employ the aRIP constants from Definition~\ref{def:LU}, Lemma~\ref{lem:arip}, and we maintain the smallest support size in $L(\cdot,n,N),U(\cdot,n,N)$.

\begin{lemma}\label{prop:SPnoisyA}
For $x\in\chi^N(k)$ and $y=Ax+e$, after iteration $l$ of SP 
\begin{equation}\label{eq:SPnoisyA}
\left\|x_{T-\tilde{T}^l}\right\|_2 \le \frac{2 U(3k,n,N)}{1-L(k,n,N)} \left(1+ \frac{U(2k,n,N)}{1-L(k,n,N)}\right)\left\|\xTTLL\right\|_2 + \frac{2\sqrt{1+U(k,n,N)}}{1-L(k,n,N)}\|e\|_2.
\end{equation}
\end{lemma}
 
\begin{lemma}\label{prop:SPnoisyB}
For $x\in\chi^N(k)$ and $y=Ax+e$, after iteration $l$ of SP 
\begin{equation}\label{eq:SPnoisyB}
\left\|\xTTL\right\|_2 \le  \left(1+ \frac{2U(3k,n,N)}{1-L(2k,n,N)}\right)\left\|x_{T-\tilde{T}^l}\right\|_2 + \frac{2}{\sqrt{1-L(2k,n,N)}}\|e\|_2.
\end{equation}
\end{lemma}

The following lemma is an adaptation of \cite[Lemma 3]{SubspacePursuit}.  By using Definition \ref{def:LU} and selecting the smallest possible support sizes for the aRIP constants, we arrive at Lemma \ref{prop:SPnoisyC}.

\begin{lemma}\label{prop:SPnoisyC}
Let $x\in\chi^N(k)$ and $y=Ax+e$ be the measurement contaminated with noise $e$.  If the Subspace Pursuit algorithm terminates after $l$ iterations, the output $\hat{x}$ approximates $x$ within the bounds
\begin{equation}\label{eq:SPnoisyC}
\|x-\hat{x}\|_2 \le \left(1+\frac{U(2k,n,N)}{1-L(k,n,N)}\right)\|x_{T-T^l}\|_2+\frac{\sqrt{1+U(k,n,N)}}{1-L(k,n,N)}\|e\|_2.
\end{equation}
\end{lemma}

Lemmas \ref{prop:SPnoisyA}--\ref{prop:SPnoisyC} combine to prove Theorem \ref{thm:SPnoisy}.

\begin{pf}[Theorem \ref{thm:SPnoisy}]
After applying Lemma \ref{prop:SPnoisyA} to Lemma \ref{prop:SPnoisyB} we bound the entries of $x$ that have not been captured by Algorithm \ref{alg:SP}, namely
\begin{equation}\label{eq:SPnoisypf1}
\|\xTTL\|_2 \le \msp \left\|\xTTLL\right\|_2 + \phi^{sp}(k,n,N)\|e\|_2
\end{equation}
where
\begin{equation}\label{eq:SPphi}
\phi^{sp}(k,n,N):=\frac{2\sqrt{1+U(k,n,N)}}{1-L(k,n,N)}\left(1+\frac{2U(3k,n,N)}{1-L(2k,n,N)}\right)+\frac{2}{\sqrt{1-L(2k,n,N)}}.
\end{equation}
Applying \eqref{eq:SPnoisypf1} iteratively, we develop a bound in terms of the norm of $x$, by observing that $\|x_{T-T^0}\|_2\le\|x\|_2$:
\begin{equation}\label{eq:SPnoisypf2}
\|\xTTL\|_2 \le \left[\msp\right]^l \|x\|_2 + \frac{\phi^{sp}(k,n,N)}{1-\msp}\|e\|_2.
\end{equation}
The factor $\frac{\phi^{sp}(k,n,N)}{1-\msp}$ amplifying $\|e\|_2$ in \eqref{eq:SPnoisypf2} is found by induction as in the proof of Theorem~\ref{thm:CSPnoisy} in Appendix~\ref{sec:ProofsCSMP}.

From Lemma \ref{prop:SPnoisyC} with $\kappa^{sp}(k,n,N)=1+\frac{U(2k,n,N)}{1-L(k,n,N)}$, we have 
\begin{equation}\label{eq:SPnoisyC2}
\|x-\hat{x}\|_2 \le \kappa^{sp}(k,n,N)\|x_{T-T^l}\|_2+\frac{\sqrt{1+U(k,n,N)}}{1-L(k,n,N)}\|e\|_2.
\end{equation}
Applying \eqref{eq:SPnoisypf2} to \eqref{eq:SPnoisyC2},
\begin{equation}\label{eq:SPnoisyC3}
\|x-\hat{x}\|_2 \le \kappa^{sp}(k,n,N)\left[\msp\right]^l \|x\|_2+\left(\kappa^{sp}(k,n,N)\frac{\phi^{sp}(k,n,N)}{1-\msp} + \frac{\sqrt{1+U(k,n,N)}}{1-L(k,n,N)}\right)\|e\|_2.
\end{equation}
From \eqref{eq:SPxi}, we verify that
\begin{equation}
\frac{\xispk}{1-\msp}=\kappa^{sp}(k,n,N)\frac{\phi^{sp}(k,n,N)}{1-\msp} + \frac{\sqrt{1+U(k,n,N)}}{1-L(k,n,N)}
\end{equation}
which completes the proof.
\end{pf}

Having established the aRIP conditions for an arbitrary measurement matrix, we again return to the Gaussian random matrix ensemble and establish the quantitative bounds for SP from Section~\ref{sec:PhaseTrans}.

\begin{pf}[Theorem~\ref{thm:SPnoisyphase}]
Let $x, y, A,$ and $e$ satisfy the hypothesis of Theorem~\ref{thm:SPnoisyphase} and select $\e>0$.
Fix $\tau<1$ and let
\begin{align*}
z(k,n,N)&=[L(k,n,N), L(2k,n,N), U(k,n,N), U(2k,n,N), U(3k,n,N)] \\
\hbox{ and }\qquad z(\d,\r)&=[L(\d,\r), L(\d,2\r), U(\d,\r), U(\d,2\r), U(\d,3\r)].
\end{align*}
Define $\mathcal{Z}=(0,\tau)^2\times(0,\infty)^3$ and define the following functions mapping $\mathcal{Z}\rightarrow\RR$:
\begin{align}
F^{sp}[z]:=F^{sp}[z_1,\dots,z_5]&= 2\frac{z_5}{1-z_1}\left(1+\frac{2z_5}{1-z_2}\right)\left(1+\frac{z_4}{1-z_1}\right), \label{eq:Fsp} \\
K[z]:=K[z_1,\dots,z_5]&= 1+\frac{z_4}{1-z_1}, \label{eq:Ksp}\\
G^{sp}[z]:=G^{sp}[z_1,\dots,z_5]&= 2 \frac{\sqrt{1+z_3}}{1-z_1}\left(1+\frac{2z_5}{1-z_2}\right)+\frac{2}{\sqrt{1-z_2}},             \\
H[z]:=H[z_1,\dots,z_5]&= \frac{\sqrt{1+z_3}}{1-z_1}.\label{eq:Hsp}
\end{align}
For each of these functions, the gradient is clearly nonnegative
componentwise on $\mathcal{Z}$, with
the first entry of each gradient strictly positive which is
sufficient to verify the hypotheses of Lemma~\ref{lem:superF} (ii).
Moreover, from \eqref{eq:SPkappak}--\eqref{eq:SPxi} and
\eqref{eq:SPkappadr}--\eqref{eq:SPxidr}, we have
\begin{align*}
\kappa^{sp}(k,n,N)\msp&=K[z(k,n,N)]F^{sp}[z(k,n,N)], \\
\kappa^{sp}(\d,\r)\mspdr&= K[z(\d,\r)]F^{sp}[z(\d,\r)], \\
\frac{\xispk}{1-\msp}&=  K[z(k,n,N)]\frac{G^{sp}[z(k,n,N)]}{1-F^{sp}[z(k,n,N)]} + H[z(k,n,N)] ,  \\
\frac{\xispdr}{1-\mspdr} &=  K[z(\d,\r)]\frac{G^{sp}[z(\d,\r)]}{1-F^{sp}[z(\d,\r)]} + H[z(\d,\r)].  
\end{align*}
Invoking Lemma~\ref{lem:superF} for each of the functions in
\eqref{eq:Fsp}--\eqref{eq:Hsp} yields that with high probability on
the draw of $A$ from a Gaussian distribution,
\begin{align}
\kappa^{sp}(k,n,N)\left[\msp\right]^l\|x\|_2 &< \kappa^{sp}(\d,(1+\e)\r)\left[\mu^{sp}(\d,(1+\e)\r)\right]^l\|x\|_2, \label{eq:spbound1} \\ \frac{\xispk}{1-\msp}\|e\|_2 &< \frac{\xi^{sp}(\d,(1+\e)\r)}{1-\mu^{sp}(\d,(1+\e)\r)}\|e\|_2. \label{eq:spbound2}
\end{align}
Combining \eqref{eq:spbound1} and \eqref{eq:spbound2} with
Theorem~\ref{thm:SPnoisy} completes the argument, recalling  that
Lemma \ref{lem:superF} applied to $F^{sp}=\mu^{sp}$ also implies that $\mspdr$ is strictly increasing in $\r$ and so Lemma \ref{lem:levelcurveALT} holds.
\end{pf}

\def\rihteps{\rho_\epsilon^{iht}(\delta)}
\def\mihteps{\mu^{iht}(\d,[1+\epsilon]\r)}

\subsection{Proofs for Iterative Hard
Thresholding}\label{sec:ProofsIHT}
In this section we first outline a proof of
Theorem~\ref{thm:IHTnoisy}, which follows similar lines to
that given by Blumensath and Davies in~\cite[Corollary 4]{BlDa08_iht}, while considering a
generalization to aRIP bounds, and also incorporating a
stepsize $\omega$. Having established this result for arbitrary
measurement matrices, we then go on to prove Theorem~\ref{thm:IHTnoisyphase}
which gives conditions for high-probability convergence of IHT in
the specific case of Gaussian random matrices. 

\begin{pf}[Theorem~\ref{thm:IHTnoisy}] 
Let $B^l=T^l\cup\mbox{supp}(x)$. Since $|B^l|\le 2k\le
3k$, we can deduce from Lemma~\ref{lem:arip} that
\begin{equation}\label{IHTcor4.2}
\|(I-\omega A_{B^l}^{\ast}A_{B^l})(x^{l-1}_{T^{l-1}}-x)_{B^l}\|_2\le\phi^{iht}(3k,n,N)\|(x^{l-1}_{T^{l-1}}-x)_{B^l}\|_2,
\end{equation}
where $\phi^{iht}(3k,n,N)$ is defined to be
$$\phi^{iht}(3k,n,N)=\max\left\{\omega\left[1+U(3k,n,N)\right]-1,1-\omega\left[1-L(3k,n,N)\right]\right\}.$$
Furthermore, we have
$$(\omega A^{\ast}_{B^l}A_{(B^{l-1}-B^l)})\subseteq(\omega
A^{\ast}_{(B^l\cup B^{l-1})}A^{\ast}_{(B^l\cup B^{l-1})}-I).$$
Since the eigenvalues of a submatrix are bounded in magnitude by the
eigenvalues of the entire matrix, and since $|B^l\cup B^{l-1}|\le 3k$,
we can again invoke Lemma~\ref{lem:arip} to obtain
\begin{equation}\label{IHTcor4.3}
\|\omega A^{\ast}_{B^l}A_{(B^{l-1}-B^l)}(x^{l-1}_{T^{l-1}}-x)_{(B^{l-1}-B^l)}\|_2\le\phi^{iht}(3k,n,N)\|(x^{l-1}_{T^{l-1}}-x)_{(B^{l-1}-B^l)}\|_2.
\end{equation}
Now we have from the proof of~\cite[Corollary 4]{BlDa08_iht} that
\begin{equation}\label{IHTcor4}
\|x^l_{T^l}-x\|_2\le
2\|(I-\omega
A^{\ast}_{B^l}A_{B^l})(x^{l-1}_{T^{l-1}}-x)_{B^l}\|_2+2\|\omega
A^{\ast}_{B^l}A_{(B^{l-1}-B^l)}(x^{l-1}_{T^{l-1}}-x)_{(B^{l-1}-B^l)}\|_2+2\|\omega
A^{\ast}_{B^l}e\|_2.
\end{equation}
Substituting (\ref{IHTcor4.2}) and (\ref{IHTcor4.3}) into
(\ref{IHTcor4}), and applying Lemma~\ref{lem:arip} to the error term,
we obtain
$$\|x^l_{T^l}-x\|_2\le
2\phi^{iht}(k,n,N)\left(\|(x^{l-1}_{T^{l-1}}-x)_{B^l}\|_2+\|(x^{l-1}_{T^{l-1}}-x)_{(B^{l-1}-B^l)}\|_2\right)+2\omega\sqrt{1+U(2k,n,N)}\|e\|_2.$$
Now $B^l$ and $(B^{l-1}-B^l)$ are disjoint, so we have
$$\|(x^{l-1}_{T^{l-1}}-x)_{B^l}\|_2+\|(x^{l-1}_{T^{l-1}}-x)_{(B^{l-1}-B^l)}\|_2\le\sqrt{2}\|(x^{l-1}_{T^{l-1}}-x)_{B^l\cup(B^{l-1}-B^l)}\|_2,$$
from which it now follows that
$$\|x^l_{T^l}-x\|_2\le\mu^{iht}(k,n,N)\|x^{l-1}_{T^{l-1}}-x\|_2+\xi^{iht}(k,n,N)\|e\|_2,$$
with $\mu^{iht}(k,n,N)$ and $\xi^{iht}(k,n,N)$ defined in \eqref{eq:IHTCk} and
\eqref{eq:IHTxi}, respectively.
Given our assumption that $\miht<1$, an induction argument analogous to the induction in the
proof of Theorem~\ref{thm:CSPnoisy} gives the stronger result 
$$\|x^l_{T^l}-x\|_2\le\left[\miht\right]^l\|x\|_2+\xiiht\left(\frac{1-\left[\miht\right]^l}{1-\miht}\right)\|e\|_2.$$
We finally note that if IHT terminates after $l$ iterations we have
$\hat{x}=x^l_{T^l}$, from which the results now follows.
\end{pf}  

Armed with the results of Section~\ref{sec:ARIP} for IHT, we return to the family of Gaussian random matrices and prove the quantitative bounds for IHT from Section~\ref{sec:PhaseTrans}.

\begin{pf}[Theorem~\ref{thm:IHTnoisyphase}]
Let $x, y, A$ and $e$ satisfy the hypothesis of Theorem~\ref{thm:IHTnoisyphase} and select $\e>0$.
Fix $\tau<1$ and let
\begin{align*}
z(k,n,N)&=[L(3k,n,N), U(2k,n,N), U(3k,n,N)] \\
\hbox{ and }\qquad z(\d,\r)&=[L(\d,3\r), U(\d,2\r), U(\d,3\r)].
\end{align*}
Define $\mathcal{Z}=(0,\tau)\times(0,\infty)^2$.  For an arbitrary weight $\w\in(0,1)$, define the functions $F^{iht}_{\w}, G^{iht}_{\w} :\mathcal{Z}\rightarrow\RR$:
\begin{align}
F^{iht}_{\w}[z]:=F^{iht}_{\w}[z_1,z_2,z_3]&= 2\sqrt{2}\max\left\{\w[1+z_3]-1,1-\w[1-z_1]\right\}, \label{eq:Fiht} \\
G^{iht}_{\w}[z]:=G^{iht}_{\w}[z_1,z_2,z_3]&= \frac{\w}{\sqrt{2}}\left(\frac{\sqrt{1+z_2}}{1-\max\left\{\w[1+z_3]-1,1-\w[1-z_1]\right\}}\right). \label{eq:Giht}
\end{align}
[Note that   $F^{iht}_{\w}[z(k,n,N)]=\mu^{iht}(k,n,N)$ and
$G^{iht}_{\w}[z(k,n,N)]=\xi^{iht}(k,n,N)/(1-\mu^{iht}(k,n,N))$ due to \eqref{eq:IHTCk} and \eqref{eq:IHTxi}.]
Clearly the functions are nondecreasing so that, with any
$t\in\mathcal{Z}$, $\left(\nabla F^{iht}_{\w}[t]\right)_i\ge0$ and
$\left(\nabla G^{iht}_{\w}[t]\right)_i\ge0$ for $i=1,2,3$; note that
$F^{iht}_{\w}[t]$ and $ G^{iht}_{\w}[t]$ have points of
nondifferentiability, but that the left and right derivatives at those points 
remain nonnegative. 
Also, and for any $v\in\mathcal{Z}$, since $t_i,v_i>0$ for each $i$, $\nabla F^{iht}_{\w}[t]\cdot v > 0$ and $\nabla G^{iht}_{\w}[t]\cdot v > 0$ as both functions clearly increase when each component of the argument increases.  Hence, $F^{iht}_{\w}$ and $G^{iht}_{\w}$ satisfy the hypotheses of Lemma~\ref{lem:superF} (i).  Therefore, for any $\w\in(0,1)$, as $(k,n,N)\rightarrow\infty$ with $\nN\rightarrow\d$, $\kn\rightarrow\r$, 
\begin{align}
\hbox{Prob}\left(F^{iht}_{\w}[z(k,n,N)]<F^{iht}_{\w}[z(\d,\r)+1c\e]\right)&\rightarrow1,\label{eq:Fiht1} \\
\hbox{Prob}\left(G^{iht}_{\w}[z(k,n,N)]<G^{iht}_{\w}[z(\d,\r)+1c\e]\right)&\rightarrow1.\label{eq:Giht1}
\end{align}

Now fix $\w^\star:=\frac{2}{2+U(\d,3\r)-L(\d,3\r)}$ and define
\begin{align}
\tilde{F}^{iht}_{\w^\star}[z]:=\tilde{F}^{iht}_{\w^\star}[z_1,z_2,z_3]&= 2\sqrt{2}\left(\frac{z_1+z_3}{2+z_3-z_1}\right),\label{eq:Fihtdr}\\
\tilde{G}^{iht}_{\w^\star}[z]:=\tilde{G}^{iht}_{\w^\star}[z_1,z_2,z_3]&= \frac{4\sqrt{1+z_2}}{2-(2\sqrt{2}-1)z_3-(2\sqrt{2}+1)z_1}.\label{eq:Gihtdr}
\end{align}
Then for any $t\in\mathcal{Z}$, $\left(\nabla \tilde{F}^{iht}_{\w^\star}[t]\right)_i>0$ for $i=1,3$ and $\left(\nabla \tilde{F}^{iht}_{\w^\star}[t]\right)_2=0$.  Likewise,  $\left(\nabla \tilde{G}^{iht}_{\w^\star}[t]\right)_i>0$ for $i=1,2,3$. Thus $\tilde{F}^{iht}_{\w^\star}$ and $\tilde{G}^{iht}_{\w^\star}$ satisfy the hypotheses of Lemma~\ref{lem:superF} (ii) and, therefore, 
\begin{align}
\tilde{F}^{iht}_{\w^\star}[z(\d,\r)+1c\e]&<\tilde{F}^{iht}_{\w^\star}[z(\d,(1+\e)\r)],\label{eq:Fiht2} \\
\tilde{G}^{iht}_{\w^\star}[z(\d,\r)+1c\e]&<\tilde{G}^{iht}_{\w^\star}[z(\d,(1+\e)\r)].\label{eq:Giht2}
\end{align}

Finally, observe that 
\begin{align}
F^{iht}_{\w^\star}[z(\d,\r)+1c\e]&=\tilde{F}^{iht}_{\w^\star}[z(\d,\r)+1c\e], \label{eq:Fihtdr1} \\
G^{iht}_{\w^\star}[z(\d,\r)+1c\e]&=\tilde{G}^{iht}_{\w^\star}[z(\d,\r)+1c\e]. \label{eq:Gihtdr1}
\end{align}
In \eqref{eq:Fiht1} and \eqref{eq:Giht1}, the weight was arbitrary; thus both statements certainly hold for the particular weight $\w^\star$.  Therefore, combining \eqref{eq:Fiht1}, \eqref{eq:Fiht2}, \eqref{eq:Fihtdr1} and combining \eqref{eq:Giht1}, \eqref{eq:Giht2}, \eqref{eq:Gihtdr1} imply that with exponentially high probability on the draw of $A$,
\begin{align}
F^{iht}_{\w^\star}[z(k,n,N)]&<\tilde{F}^{iht}_{\w^\star}[z(\d,(1+\e)\r)],\label{eq:Fiht3} \\
G^{iht}_{\w^\star}[z(k,n,N)]&<\tilde{G}^{iht}_{\w^\star}[z(\d,(1+\e)\r)].\label{eq:Giht3}
\end{align}
Therefore, with the weight $\w^\star$, there is an exponentially high
probability on the draw of $A$ from a Gaussian distribution that
\begin{align}
\miht = F^{iht}_{\w^\star}[z(k,n,N)]&<\tilde{F}^{iht}_{\w^\star}[z(\d,(1+\e)\r)] = \mu^{iht}(\d,(1+\e)\r),\label{eq:Fiht4} \\
\frac{\xiihtk}{1-\miht} = G^{iht}_{\w^\star}[z(k,n,N)]&<\tilde{G}^{iht}_{\w^\star}[z(\d,(1+\e)\r)] = \frac{\xi^{iht}(\d,(1+\e)\r)}{1-\mu^{iht}(\d,(1+\e)\r)},\label{eq:Giht4}
\end{align}
where we also employed \eqref{eq:IHTCk}, \eqref{eq:IHTxi} with $\w=\w^*$,
and \eqref{eq:IHTmudr}, \eqref{eq:IHTxidr}. 
The result follows by invoking Theorem~\ref{thm:IHTnoisy} and applying
\eqref{eq:Fiht4} and \eqref{eq:Giht4};
recall also that Lemma \ref{lem:levelcurveALT} holds since
$\mu^{iht}(\d,\r)=\tilde{F}^{iht}_{\w^\star}(z(\d,\r))$ is implied to
be strictly increasing in $\r$ by Lemma \ref{lem:superF} (ii).
\end{pf}

\bibliographystyle{plain}
\bibliography{GreedyPT_arXiv}

\end{document}